\documentclass[usenatbib]{mn2e}
\usepackage{graphicx}
\usepackage{ifthen}
\usepackage{url}

\def\ltsima{$\; \buildrel < \over \sim \;$}
\def\lta{\lower.5ex\hbox{\ltsima}}
\def\gtsima{$\; \buildrel > \over \sim \;$}
\def\simgt{\lower.5ex\hbox{\gtsima}}
\def\arcsec{\mathop{\rm arcsec}\nolimits} 

\def\kms{{\rm\,km\,s^{-1}}}

\def\kpc{{\rm\,kpc}}

\def\AA{$\; \buildrel \circ \over {\rm A}$}

\def\s{\ifmmode \widetilde \else \~\fi}
\def\={\overline}

\def\spose#1{\hbox to 0pt{#1\hss}}

\def\lta{\mathrel{\spose{\lower 3pt\hbox{$\mathchar"218$}}
     \raise 2.0pt\hbox{$\mathchar"13C$}}}
\def\gta{\mathrel{\spose{\lower 3pt\hbox{$\mathchar"218$}}
     \raise 2.0pt\hbox{$\mathchar"13E$}}}
\def\Dt{\spose{\raise 1.5ex\hbox{\hskip3pt$\mathchar"201$}}}    
\def\dt{\spose{\raise 1.0ex\hbox{\hskip2pt$\mathchar"201$}}}    

\def\dotsfill{\leaders\hbox to 1em{\hss.\hss}\hfill}

\def\FeH{{\rm[Fe/H]}}

\def\andxi{And\,XI}
\def\andxii{And\,XII}
\def\andxiii{And\,XIII}
\def\vhel{v$_{\rm hel}$}
\def\streama{Stream\,`A'}
\def\streamb{Stream\,`B'}
\def\streamc{Stream\,`C'}
\def\streamcr{Stream\,`Cr'}
\def\streamcp{Stream\,`Cp'}
\def\streamd{Stream\,`D'}
\def\ec4{EC4}

\loadboldmathitalic 
\title[Kinematics of M31 stellar streams]{
The kinematic footprints of five stellar streams  in Andromeda's halo$^{1}$
}
\author[S.\ C.\ Chapman et al.] 
{S.\ C.\ Chapman$^{2,3,4}$, R.\ Ibata$^{5}$,  M.\ Irwin$^{2}$, A.\ Koch$^{6}$,  B.\ Letarte$^7$, N.\ Martin$^{8}$,
\newauthor  
M.\ Collins$^2$,  
G.\ F.\ Lewis$^{9}$,  A.\ McConnachie$^{3}$, J.\ Pe\~narrubia$^{3}$, 
\newauthor R.\ M.\ Rich$^{6}$,  D.\ Trethewey$^{2}$,   
A.\ M.\ N.\ Ferguson$^{10}$, A.\ Huxor$^{10}$, N.\ Tanvir$^{11}$\\
$^{2}$ Institute of Astronomy, Madingley Road, Cambridge, CB3 0HA, U.K.\\
$^{3}$ Department of Physics and Astronomy, University of Victoria, Victoria, B.C., V8P 1A1, Canada\\
$^{4}$ Canadian Space Agency, Space Science Fellow\\
$^{5}$ Observatoire de Strasbourg, 11, rue de l'Universit\'e, F-67000, Strasbourg, France\\
$^{6}$ Department of Physics and Astronomy, University of California at Los Angeles\\ 
$^{7}$ California Institute of Technology, Pasadena, CA, 91125, USA\\
$^{8}$ Max-Planck-Institut f\"ur Astronomie, K\"onigstuhl 17, D-69117 
Heidelberg, Germany\\
$^{9}$ Institute of Astronomy, School of Physics, A29, University of Sydney, NSW 2006, Australia\\
$^{10}$ Institute for Astronomy, University of Edinburgh, Royal Observatory, Blackford Hill, Edinburgh, UK EH9~3HJ\\
$^{11}$ Department of Physics \& Astronomy, University of Leicester, Leicester, LE17RH, UK\\ 
}

\date{\today}
\begin{document} 
\maketitle 

\begin{abstract} 
We present a spectroscopic analysis of five stellar streams (`A', `B', `Cr', `Cp' and `D') as well as  the extended star cluster, \ec4, which lies within \streamc, all discovered in the halo of M31 from our CFHT/MegaCam survey.
These spectroscopic results were initially serendipitous, making use of our existing 
observations from the DEep Imaging Multi-Object Spectrograph mounted on
the Keck~II telescope,
and thereby emphasizing the ubiquity of tidal streams that account for $\sim$70\% of the M31 halo stars  
in the targeted fields. 
Subsequent spectroscopy was then procured in \streamc\ and \streamd\ to trace the velocity gradient along the streams.
Nine metal-rich ([Fe/H]$\sim$-0.7) stars at $v_{\rm hel}=-349.5$~km/s, $\sigma_{v, corr}\sim5.1\pm2.5$\,km/s
are proposed as a serendipitous detection of \streamcr, with followup kinematic identification at
a further point along the stream.
Six metal-poor ([Fe/H]$\sim$-1.3) stars confined to a narrow, 15~km/s velocity bin
centered at $v_{\rm hel}=-285.6$~km/s,   $\sigma_{v,corr}=4.3^{+1.7}_{-1.4}$~km/s 
represent a kinematic detection of \streamcp, again with followup kinematic identification
further along the stream. 
For the cluster \ec4, candidate member stars 
with average [Fe/H]$\sim$-1.4 ([Fe/H]$_{spec}$=-1.6), are found at 
$v_{\rm hel}=-285$~km/s  suggesting it could be related to \streamcp. 
No similarly obvious cold kinematic candidate is found for \streamd, although candidates are proposed in both of two spectroscopic pointings along the stream (both at $\sim-400$km/s). 
Spectroscopy near the edge of \streamb\ suggests a likely kinematic detection at 
$v_{\rm hel}\sim-330$~km/s, $\sigma_{v, corr}\sim6.9$\,km/s, while a candidate kinematic detection of \streama\ is found (plausibly associated to M33 rather than M31) with $v_{\rm hel}\sim-170$~km/s, $\sigma_{v, corr}=12.5$\,km/s.
The low dispersion of the streams in kinematics, physical thickness, and metallicity makes it hard to reconcile with a scenario whereby these stream structures as an ensemble are related to the giant southern stream. 
We conclude that the M31 stellar halo is largely made up of multiple kinematically cold streams.
\end{abstract}
 

\section{Introduction}
\footnotetext[1]{The data presented herein were obtained at the W.M. Keck Observatory, which is operated as a scientific
partnership among the California Institute of Technology, the University of California and the National Aeronautics and
Space Administration. The Observatory was made possible by the generous financial support of the W.M. Keck Foundation.}



Stellar streams represent the visible debris of small galaxies being
cannibalized by large galaxies, memorials to the merging process
by which the halos of galaxies are built up. 
The best known examples of streams in the Milky Way (MW) 
have recently been mapped
far more extensively by the SDSS-DR5 by Belokurov et al.\ (2006, 2007): 
the tidally stripped stars and globular clusters associated with the 
Sagittarius 
dwarf spheroidal and the Low Latitude stream, along with a 
newly discovered ``Orphan Stream'' so named for its lack of obvious progenitor.
In M31, thanks to our ability to efficiently map vast regions of the halo, the number of discovered giant streams already 
outnumbers that of the MW (Ibata et al.\ 2007). 
However, of the eight stellar streams that have been identified
in the halo of M31, only one has 
plausibly been identified to a dwarf satellite:
the loop connecting to NGC~205 (McConnachie et al.\ 2005).
This suggests that the other 
streams could represent an additional seven
`uncataloged' satellites, 
although some of the streams might be produced by a common progenitor, as suggested by models of Fardal et al.\ (2007, 2008) for the M31 Giant Southern Stream, or in a similar 
fashion to the Sgr dwarf and its numerous wraps around the Milky Way.
It is important to characterize their orbits, metallicities and masses,
to understand what their progenitors must have been.


The existence of stellar streams tells us that a progenitor galaxy 
has undergone significant mass loss.
This is due to a combination of its orbit and its phase of evolution -- the amount of dark matter mass the galaxy has lost 
(Pe\~narrubia et al.\ 2008a,b).
%
Satellites on circular orbits are harder
to disrupt, but if they are massive enough (i.e.\ of the order of the LMC),
dynamical friction will bring them close to the host galaxy centre, where
the interactions with the disc will lead to their tidal disruption -- e.g., Pe\~narrubia et al.\ 2007).
In addition, to form a stream one has to remove most of the dark matter halo
($\sim$90--99\%). The most important parameter that controls the mass loss rate
of a dSph is the pericentre distance (the orbital eccentricity is of
second order).
On the other hand dwarfs
with highly elliptical orbits spend a lot of time near apocentre where they
are unlikely to be disrupted by the host.
It is therefore not immediately obvious that streams represent preferred types of orbits
on average.
However, a number of theoretical studies have shown that significant information about the orbital properties of the progenitor galaxy can be derived from the streams (e.g, Ghigna et al.\ 1998; 
Helmi et al.\ 1999a).

Streams can be much more informative to study than dwarfs because
their orbits can be directly traced and constrained.
Fellhauer et al.\ (2006) were able to accurately constrain the shape of the Galactic potential 
through the bifurcation of Sagittarius streams in Belokurov et al.\ (2006).
Understanding the range of orbits of satellites to large galaxies will
help us to understand how the halos of these galaxies formed.
This is especially interesting in light of M31's huge stellar halo reflecting the
dark matter dominated halo out to $\simgt$150~kpc  (e.g., Irwin et al.\ 2005,
Gilbert et al.\ 2006,  Ibata et al.\ 2007).
However streams can also be much harder to analyse observationally:
the distances are problematic, there's a much lower spatial density and
they have a larger extent so that observational sampling is not trivial.
There is also the difficulty to infer the membership of
different stream pieces, especially if we
expect different chemical signatures due to metallicity
gradients in the progenitor system (e.g., Ibata et al.\ 2007).


Streamy/blobby structures 
 are individually interesting and constraining for the halo formation.
They can represent the only traceable product of long disintegrated
progenitors, yet retain a coherent body for statistical analysis.
Streams can provide
important clues on the structure of the progenitors (e.g.\ metallicity
gradients, mass to light -- M/L) as well as on the shape of the host dark matter halo (e.g.\
prolate versus oblate)
(Martinez-Delgado et al.\ 2008).
Future study of these structures will be able to put them in a
much better "near-field cosmology" context, eventually understanding their
ages and chemical histories.
However it is important to uncover and study them now, even in limited capacities necessitated by 
the small numbers of spectroscopically identifiable stars and HR-diagram depths, so we can build our
models on the most complete context.

We have initiated a spectroscopic survey of the new streams found
in M31's halo
using the DEep Imaging Multi-Object Spectrograph on Keck~II to derive 
radial velocities and metallicities of red giant branch (RGB) stars.
In this contribution, we discuss spectroscopic pointings in 
each of streams `C' and  `D' which we obtained by serendipity, since the
spectroscopy was taken prior to knowledge of the photometrically 
discovered streams, as well as followup spectroscopic pointings in both of these streams.
We also analyze spectroscopic data from Koch et al.\ (2008) lying within the Ibata et al.\ (2007) streams `A' and `B'.



\section{Observations and analysis}

\begin{figure*}
\label{orient}
\begin{center}
\includegraphics[angle=270,width=0.80\hsize]{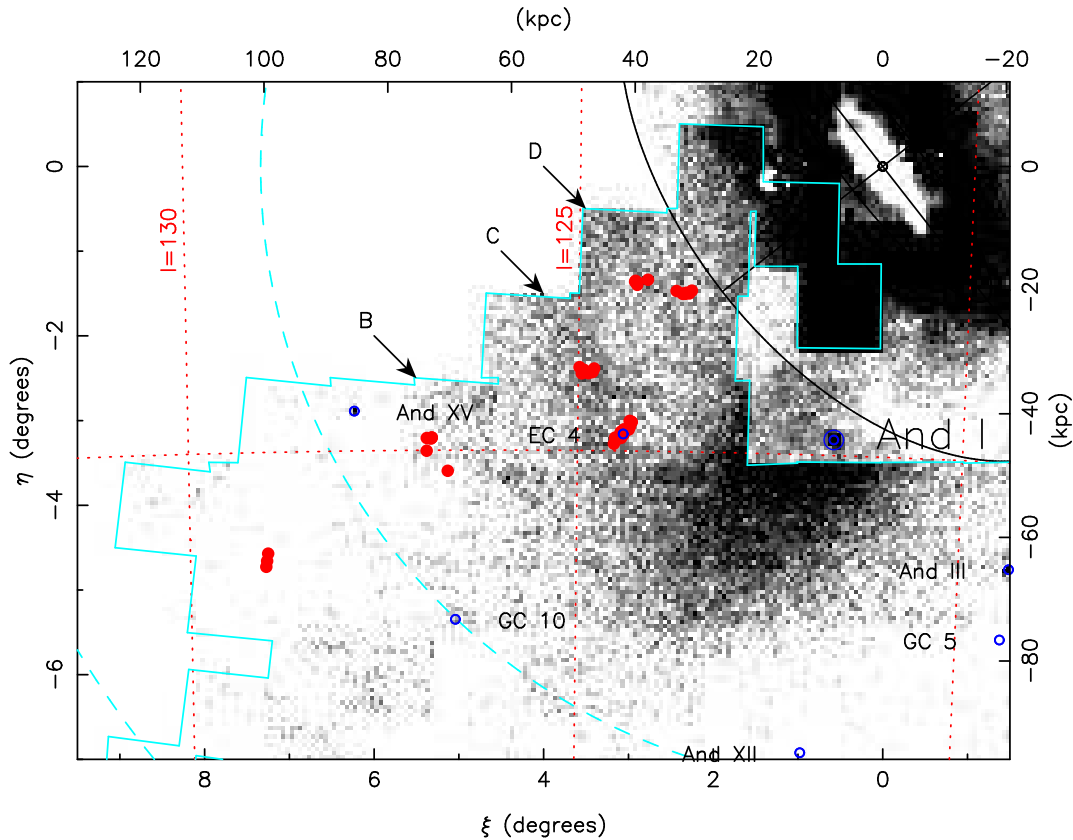}
\caption{
The locations of the spectroscopic measurements (red circles) are overlaid on the imaging data from the INT and CFHT telescopes presented in Ibata et al. (2007). Streams `B', `C' and `D' intersect the minor axis of M31 approximately perpendicularly at $\sim 80\kpc$, $\sim 60\kpc$ and $\sim 35\kpc$, respectively. The irregular turquoise  line demarks the CFHT survey region. The scale of the diagram is shown by the circle segment (dashed line - marking a projected radius of $100\kpc$), as well as by the ellipse segment (continuous line - showing a $50\kpc$ ellipse of axis ratio 0.6).
Two of the newly discovered dwarf spheroidals from our survey are visibile in the same region, 
And~XII (Martin et al.\ 2006, Chapman et al.\ 2007), and And~XV (Ibata et al.\ 2007, Letarte et al.\ 2008).
}
\end{center}
\end{figure*}


The spectroscopic fields along the M31 minor axis discussed in this paper are highlighted in 
Figure~1, two lying at $\sim$35~kpc, three lying at $\sim$60~kpc, one at $\sim$80~kpc, and one at $\sim$120~kpc
projected from the center of M31. The fields cover the four stellar streams presented in the Ibata et al.\ (2007) M31 extended halo analysis,
called \streamd, `C', `B', and `A' respectively. 
In these M31 halo images, it can be seen that \streamc\ has significantly different morphology as a function of metallicity, a more metal-rich component dominating the structure, with a more irregular shaped metal-poor component to the east. While it was not proposed initially, our evidence in this work suggests the two structures may be distinct systems, and we refer to these as two separate streams, \streamcr\ for the metal-rich component and \streamcp\ for the metal-poor component (this issue is explored in detail in \S~3.3 and Figs.~10 \& 11).

Multi-object spectroscopic observations with the Keck-II telescope and 
the DEep Imaging Multi-Object Spectrograph  -- DEIMOS (Davis et al.\ 2003) 
were obtained in photometric conditions with  $\sim0.8''$ seeing in Sept.\ 2004 and 2005.
Target stars were chosen by colour/magnitude selection as described
in Ibata et al.\ (2005), first selecting likely RGB stars in M31 over all
metallicities, and filling space with any other stellar objects in the field.
Two spectroscopic masks (F25 and F26 from the table in
Chapman et al.\ 2006) targeted the
field of an extended cluster \ec4\ (Mackey et al.\ 2006), 
which were found after the fact to be spanning \streamcr\ and \streamcp.
A combined total of 212 independent stars in both masks
were observed in standard DEIMOS slit-mask mode (Davis et al.\ 2003) 
using the high resolution 1200 line/mm grating, and 1$''$ width slitlets.
Ten of these target stars were specifically selected from HST photometry of
\ec4\ to lie within the cluster.
Our instrumental setting 
covered the observed wavelength range from $\sim0.70$--0.98\,$\mu$m.
Exposure time was 60 min, split into 20-min integrations. 
The DEIMOS-DEEP2 pipeline (Newman et al.\ 2004) 
designed to reduce data of this type accomplishes tasks of
debiassing, flat-fielding, extracting, wavelength-calibrating and
sky-subtracting the spectra.
The same settings were used to target a halo field which was
found after the fact to lie in \streamd.
89 stars were observed in this mask (F7).

The Ibata et al.\ (2007) imaging discovery of the new streams, coupled with obvious kinematic detection of the \streamcr\ and \streamcp\ in our existing spectroscopic observations (described in subsequent sections) 
prompted the followup study of these structures. On October 8, 
2007, additional DEIMOS masks were obtained further along the \streamc\ and \streamd, as identified in Fig.~1 and Table~1.
These observations were obtained under $\sim1\arcsec$\ seeing, and cloudy conditions.
For the field F36, 
we obtained 3$\times$20\,min integrations on a mask with 113 targeted stars, while field F37 
was observed for 4$\times$20\,min integrations with 138 targeted stars.
These observations used the lower resolution 600l/mm grating to achieve a higher S/N in the continuum for the fainter stars, and resulting in a resolution of $\sim$3\AA\ estimated from the width of sky lines.

The  radial velocities  of the  stars in all these fields were
then  measured with  respect  to spectra  of  standard stars  observed
during the observing runs. By fitting the peak of the cross-correlation
function, an estimate of the radial velocity accuracy was obtained for
each  radial velocity  measurement.  The accuracy  of  these data,
as estimated from the Calcium Triplet (CaT) cross-correlation,
varies with magnitude, having uncertainties  of
$<10\kms$ for most of the stars. The CMDs, velocity errors, velocity histograms, and metallicities for these fields are shown in Figures~2 \& 3 for \streamc, and Figures~4 \& 5 for \streamd.
Spectroscopic metallicities quoted in these tables are calculated from the equivalent widths of the Ca{\sc II} triplet lines, as described in Ibata et al.\ (2005).

Finally, the two furthest streams along the minor axis from Ibata et al.\ (2007),  \streama\ and \streamb\ lying at 120~kpc and 80~kpc respectively, have serendipitous DEIMOS spectroscopic pointings lying in their edge regions from
Gilbert  et al.\ (2006) and Koch et al.\ (2008) (fields M8 and M6 respectively).
Reduction and analysis of these two fields is detailed in Koch et al.\ (2008).
Figures~6 \& 7 show the CMDs and velocity/metallicity distributions for \streamb\ \& \streama\ respectively. 

We address Galactic contamination to our spectroscopically identified stars in a manner identical to Koch et al.\ (2008), using a combination of $V-I$ radial velocity and the equivalent width (EW) of  {Na\sc i}$_{\lambda8183,8195}$ which is
sensitive to surface gravity, and is accordingly very weak in M31
RGB star spectra, but can be strong in Galactic dwarfs (Schiavon et al.\ 1997). 
At velocities v$_{\rm hel}<-150$~km/s, very few RGB candidates show any
significant {Na\sc i}$_{\lambda8183,8195}$ absorption lines,
whereas stars with $-150$ to $0$~km/s
velocity show strong {Na\sc i} absorption on average, consistent with the findings of
Guhathakurta et al.\ (2006), Chapman et al.\ (2006), Gilbert et al.\ (2006) and Koch et al.\ (2008).
For this study, we impose the additional constraint of removing all stars from the halo sample with v$_{\rm hel}>-150$~km/s, and we remove any stars from our sample which have a summed EW({Na\sc i}$_{\lambda8183,8195}$)$>0.8$   
in the velocity range v$_{\rm hel}<-150$~km/s.

The properties of all candidate M31 halo (and stream) stars in these fields are listed in Tables~2--7, including coordinates, velocities, spectroscopic and photometric metallicities, and $V$,$I$ photometry.

\begin{figure*}
\label{streamc}
\begin{center}
\includegraphics[angle=0,width=0.45\hsize]{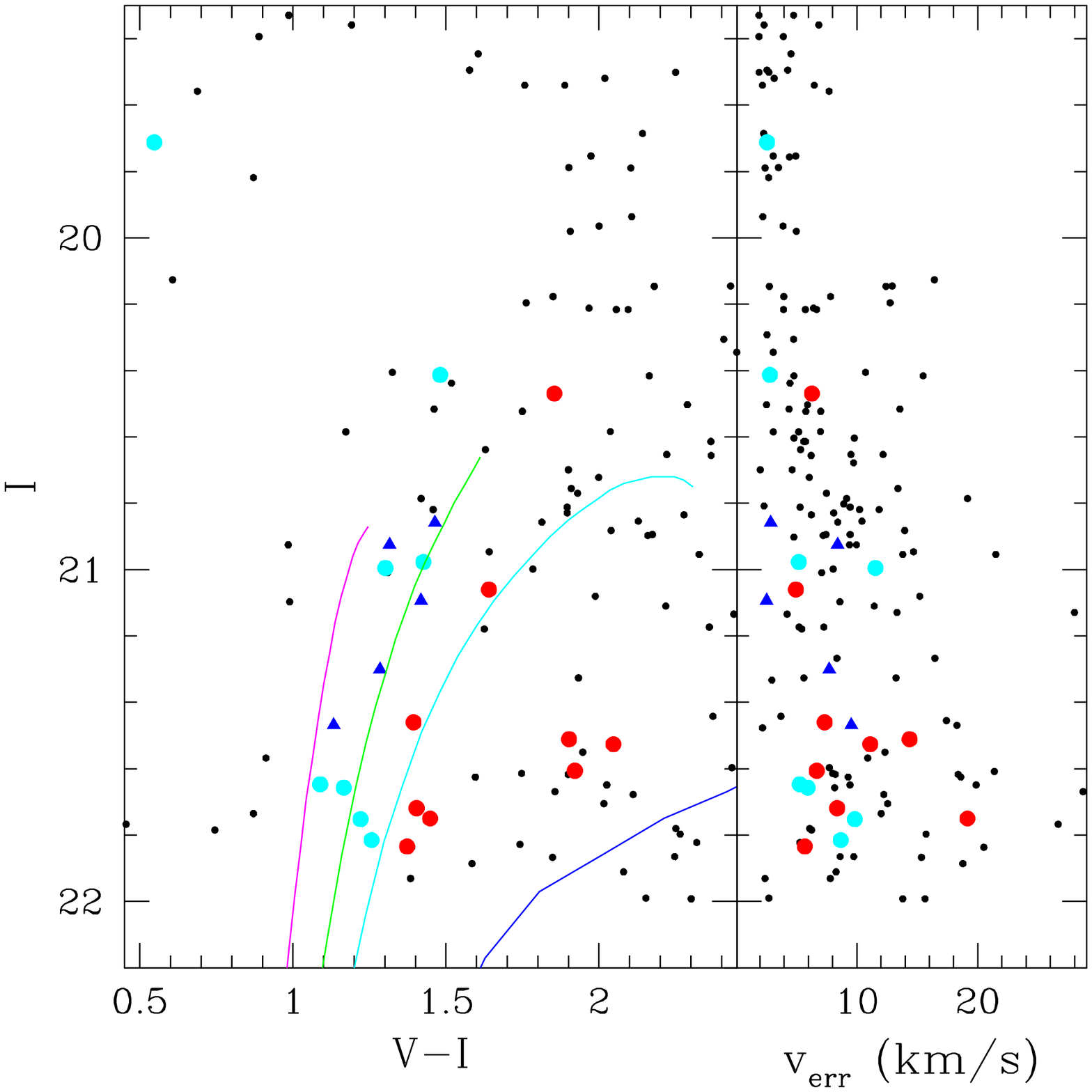}
\includegraphics[angle=0,width=0.45\hsize]{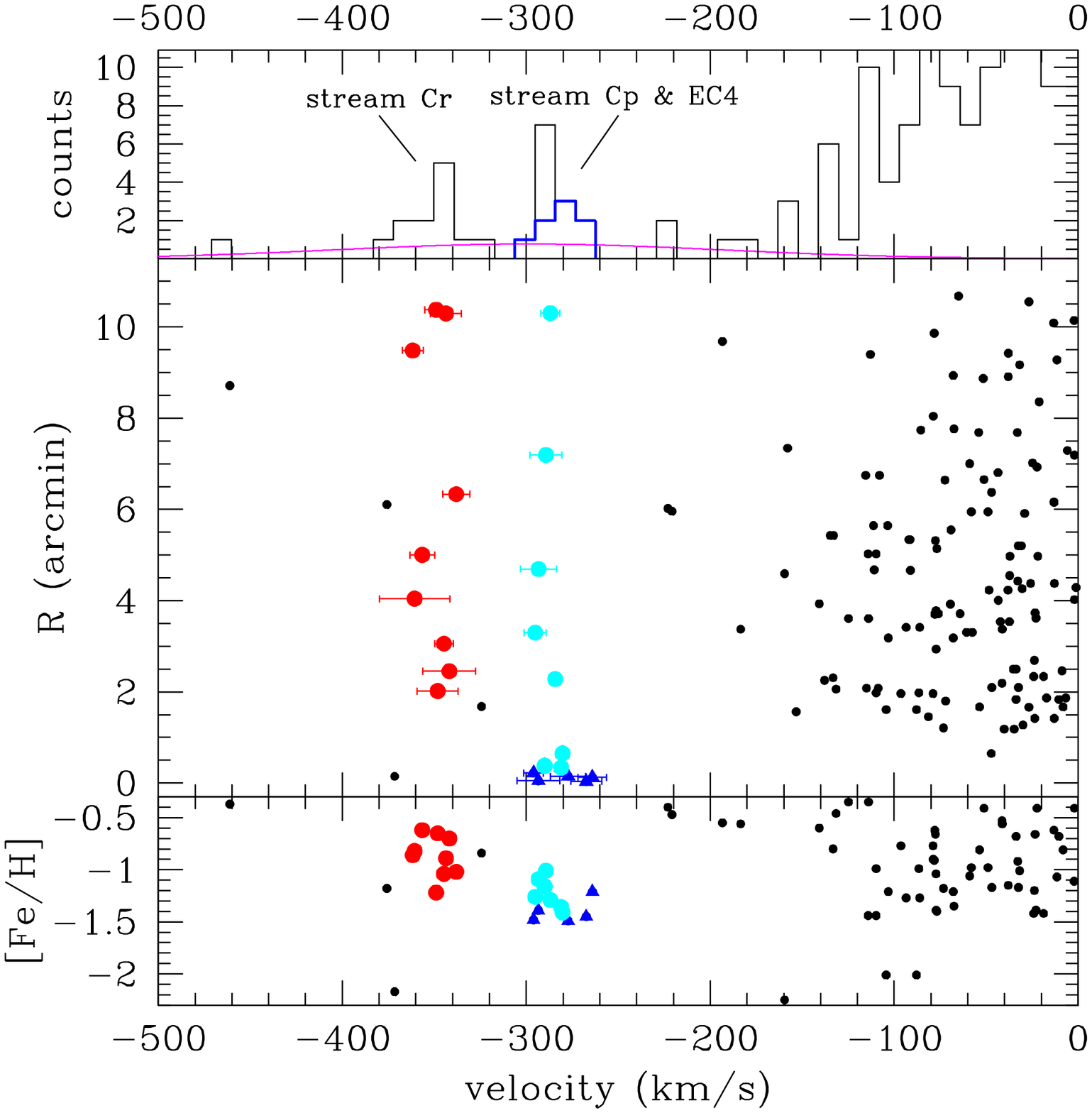}
\caption{
{\bf Left panel:}
 CFHT-MegaCam color-magnitude diagram and radial velocity uncertainties 
of the observed stars in the \streamcr/\streamcp\ fields.
Stars likely belonging to \streamcr\ (red), \streamcp\ (cyan), and \ec4\ (blue) are highlighted.
The bright and very blue star belonging to \streamcp\ is unusual for M31 and its properties are described in the text.
The fiducial RGBs correspond to, from left to right, NGC~6397, NGC~1851, 47~Tuc, NGC~6553 which have metallicity of [Fe/H]$=$-1.91, -1.29, -0.71, and -0.2,
respectively. These fiducials have been shifted to the
average distance modulus of EC4, 24.47 (785~kpc -- Mackey et al.\ 2006).
{\bf Right panel:}
The velocities of observed stars in the F25/F26 fields are shown
as a histogram, with EC4 member stars highlighted as a heavy histogram.
The stellar halo velocity dispersion ($\sigma_v$=125~km/s) 
from Chapman et al.\ (2006) is 
shown normalized to the expected 9 halo stars at this position from 
Ibata et al.\ (2007).
To differentiate EC4 stars from the field, we additionally plot
the velocities against their radius from the \ec4\ center (Table~1),
referencing the symbols to the CMD plot.
Photometrically derived [Fe/H] is shown as a function
of radial velocity,  again referenced in symbol type to the CMD plot.
}
\end{center}
\end{figure*}

\begin{figure*}
\label{streamc}
\begin{center}
\includegraphics[angle=0,width=0.45\hsize]{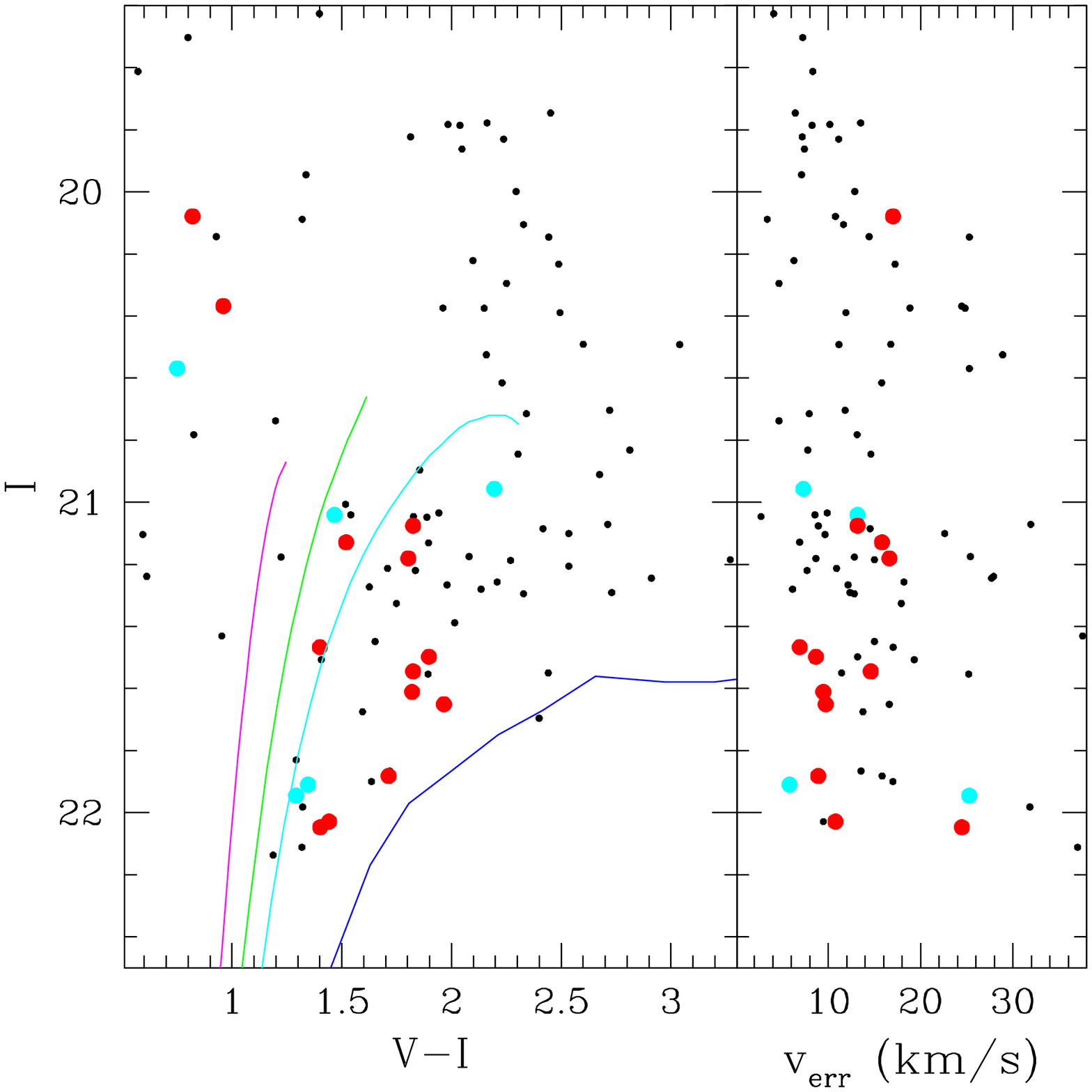}
\includegraphics[angle=0,width=0.45\hsize]{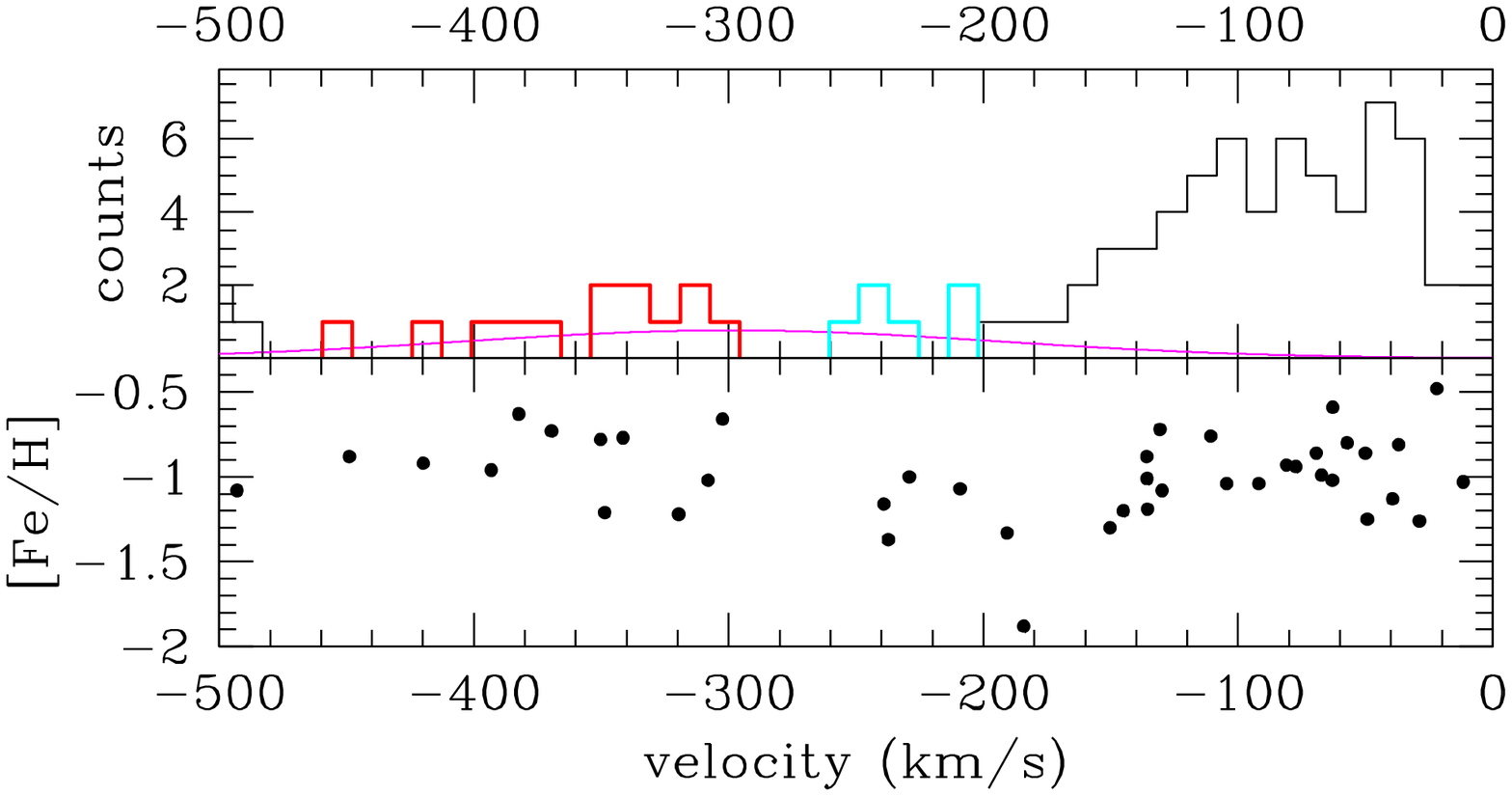}
\caption{
The same as for Fig.~2 for a field further along the \streamc\ structure (F36).
No similarly obvious kinematic peaks are detected as in Fig.~2. In this Figure only we highlight stars potentially  belonging to \streamcr\ (red) and \streamcp (cyan) entirely by their velocity ranges consistent with the spikes found in Fig.~2. Clusters of metal-poor and metal-rich stars are indeed
found in these velocity ranges consistent with belonging to stream`Cp' and stream`Cr' respectively, although it is difficult to separate them from field/halo stars.
%
}
\end{center}
\end{figure*}

\subsection{Observations of the cluster, EC4}

The two minor-axis stellar halo fields, F25/F26,
lying serendipitously in \streamc\  were observed with the additional goal of constraining
the kinematics of an ``extended star cluster'', \ec4.
These extended, luminous objects in the outskirts of M31 represent a 
population with
$\sim15$ -- 60~kpc projected radii, large half-light radii for GCs 
with luminosities near the peak of the
GC luminosity function (Huxor et al.\ 2005, 2008). As such, they are dissimilar to any other known
clusters in the Milky Way or M31, and begin to fill in the gap in parameter
space between classical globular clusters and dwarf spheroidal galaxies.
The ``faint fuzzies'' discovered in NGC~1023 (Larsen \& Brodie 2000, Brodie \& Larsen 2002),
and the similarly diffuse objects in the ACS Virgo Cluster survey (Peng et al.\ 2006),
may represent a similar class of cluster.
\ec4\ was discovered within the 
CFHT-MegaCam survey (Huxor et al.\ 2008) at a projected radius of
60~kpc (coordinates in Table 1). 

Ten stars were selected to lie within \ec4\ by their CMD colours.
Of these ten, one 
was subsequently identified as a galaxy in the HST/ACS image (Mackey et al.\ 2006), another 
lies well off the cluster RGB, likely due to contamination in the ground-based 
CFHTmegacam imaging affecting the colour measurements, while a third lies at $\sim$5 core radii and has a low probability of being associated to the cluster. The remaining seven stars are
candidate \ec4\ members, lying within 3 core radii of the cluster centre
and falling along the top of the narrow RGB from the HST imagery 
(Collins et al.\ in preparation).

In Fig.~2, the MegaCam photometry is shown for EC4 stars for consistency with the stream data.
However comparison with the very narrow RGB from HST photometry in Mackey et al.\ (2006) shows that crowding affects the ground-based accuracy, since the scatter on the Megacam CMD is much larger than the difference in photometric errors (HST to Megacam) would warrant.
%
In analyzing the DEIMOS spectroscopy, one targeted star lay exactly on the 
\ec4\ CMD, but had a discrepant velocity from the others. Closer
examination of the spectrum 
revealed good detections of the first and second CaT lines with a velocity of -277.3km/s,
whereas the automated software pipeline derived a cross-correlation fit to larger skyline residuals.
We include this star in the catalog as a viable member of \ec4.

\begin{figure*}
\label{streame}
\begin{center}
\includegraphics[angle=0,width=0.49\hsize]{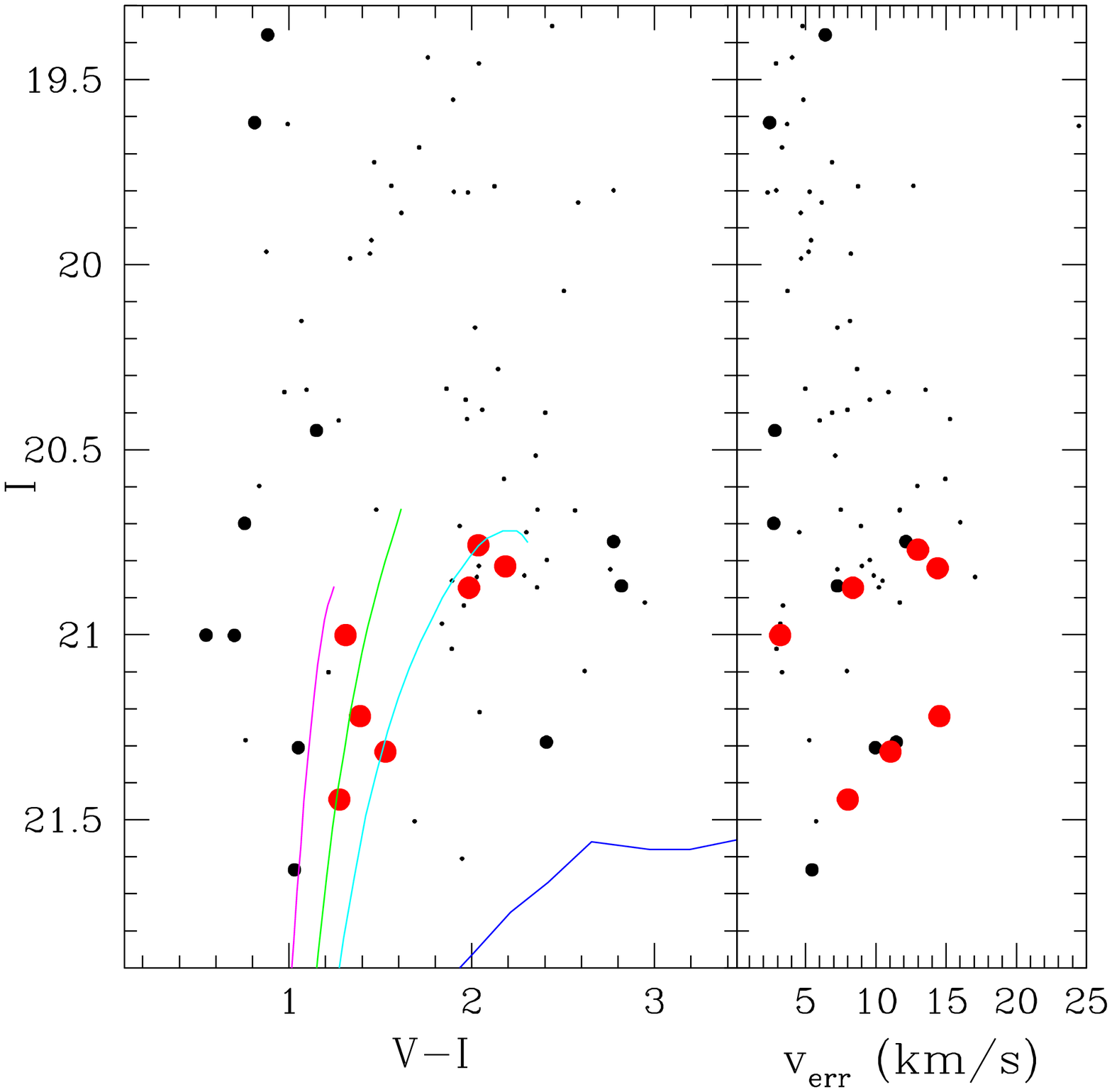}
\includegraphics[angle=0,width=0.49\hsize]{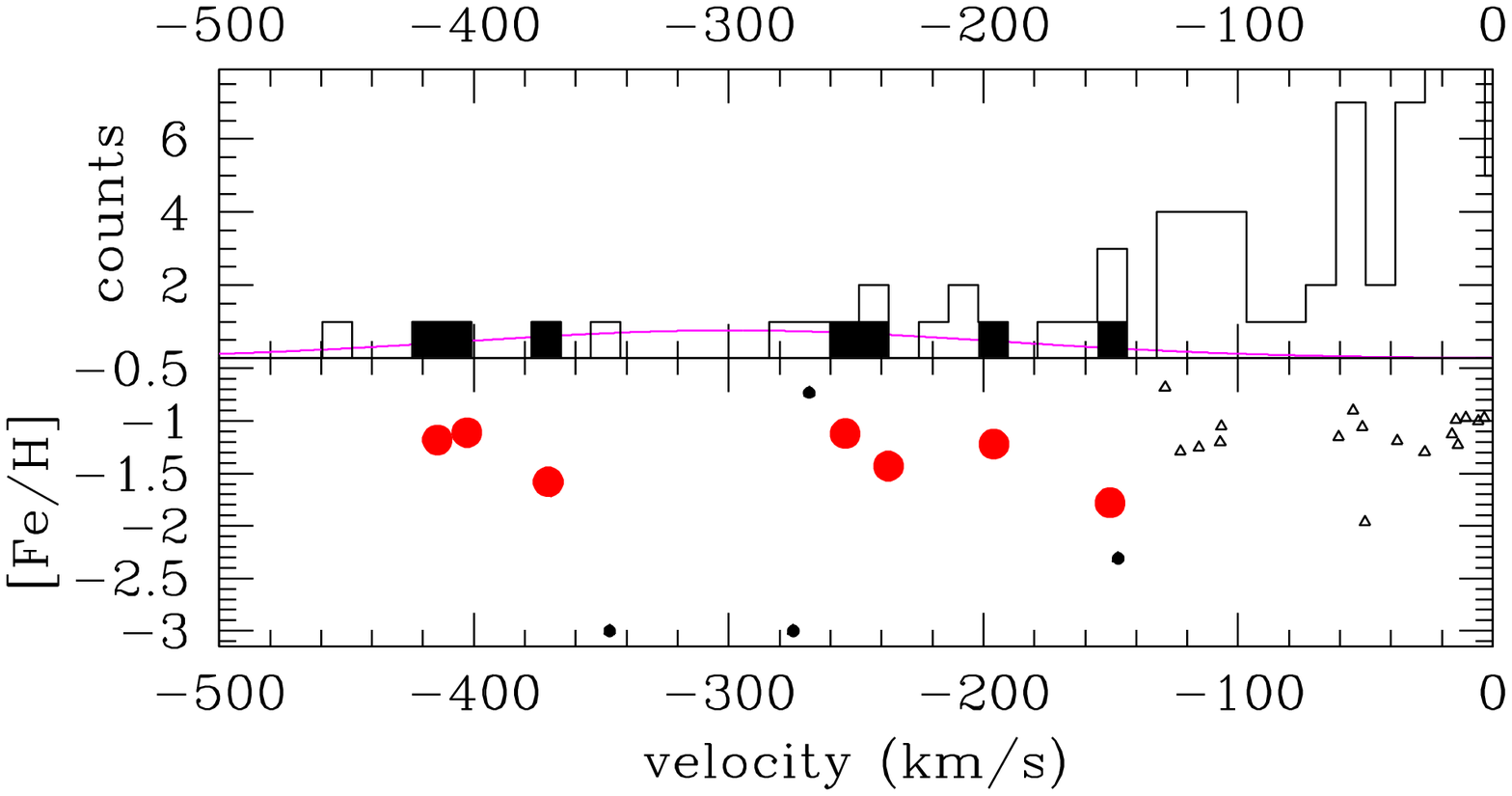}
\caption{
{\bf left panel:}
CFHT-megacam color-magnitude diagram and radial velocity uncertainties
of the observed stars in the \streamd\ field (F7).
The fiducial RGBs are as in Fig.~2, shifted to the same
average distance modulus of M31, 24.47, 785~kpc (McConnachie et al.\ 2005).
Large symbols represent stars unlikely to be contaminated by foreground
Milky Way ($v_{hel}<-150$~km/s).
Symbols are further highlighted which have photometric metallicities
consistent with the [Fe/H] distribution of \streamd\ in Ibata et al.\ (2007): $-1.7 < [Fe/H] < -0.7$.
{\bf right panel:}
The velocities of observed stars in the \streamd\ field are shown
as a histogram, with possible \streamd\ stars from the left panel 
highlighted as a filled histogram.
Photometrically derived [Fe/H] is shown as a function
of radial velocity, and referenced in symbol type to the CMD plot.
}
\end{center}
\end{figure*}

\begin{figure*}
\label{streamc}
\begin{center}
\includegraphics[angle=0,width=0.49\hsize]{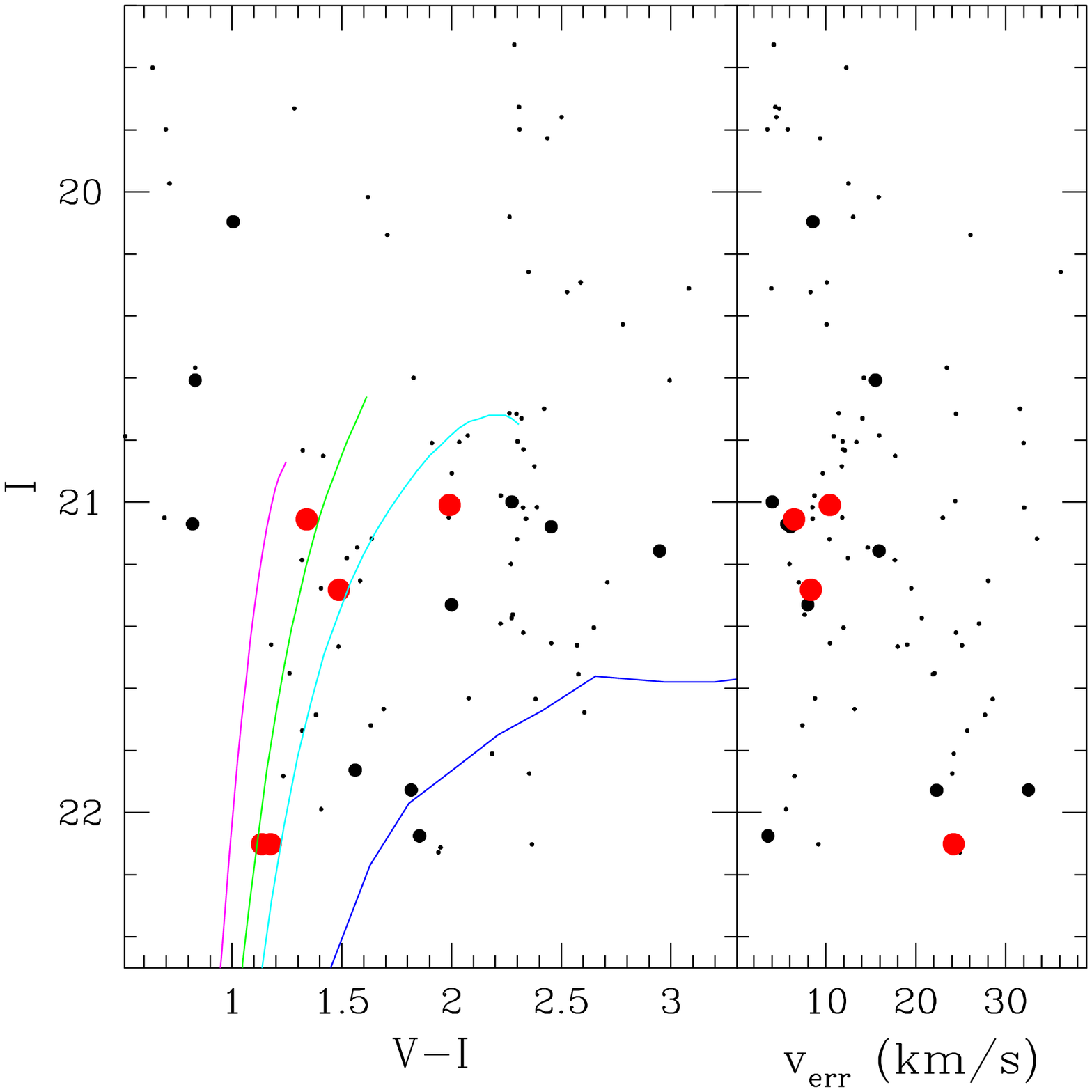}
\includegraphics[angle=0,width=0.49\hsize]{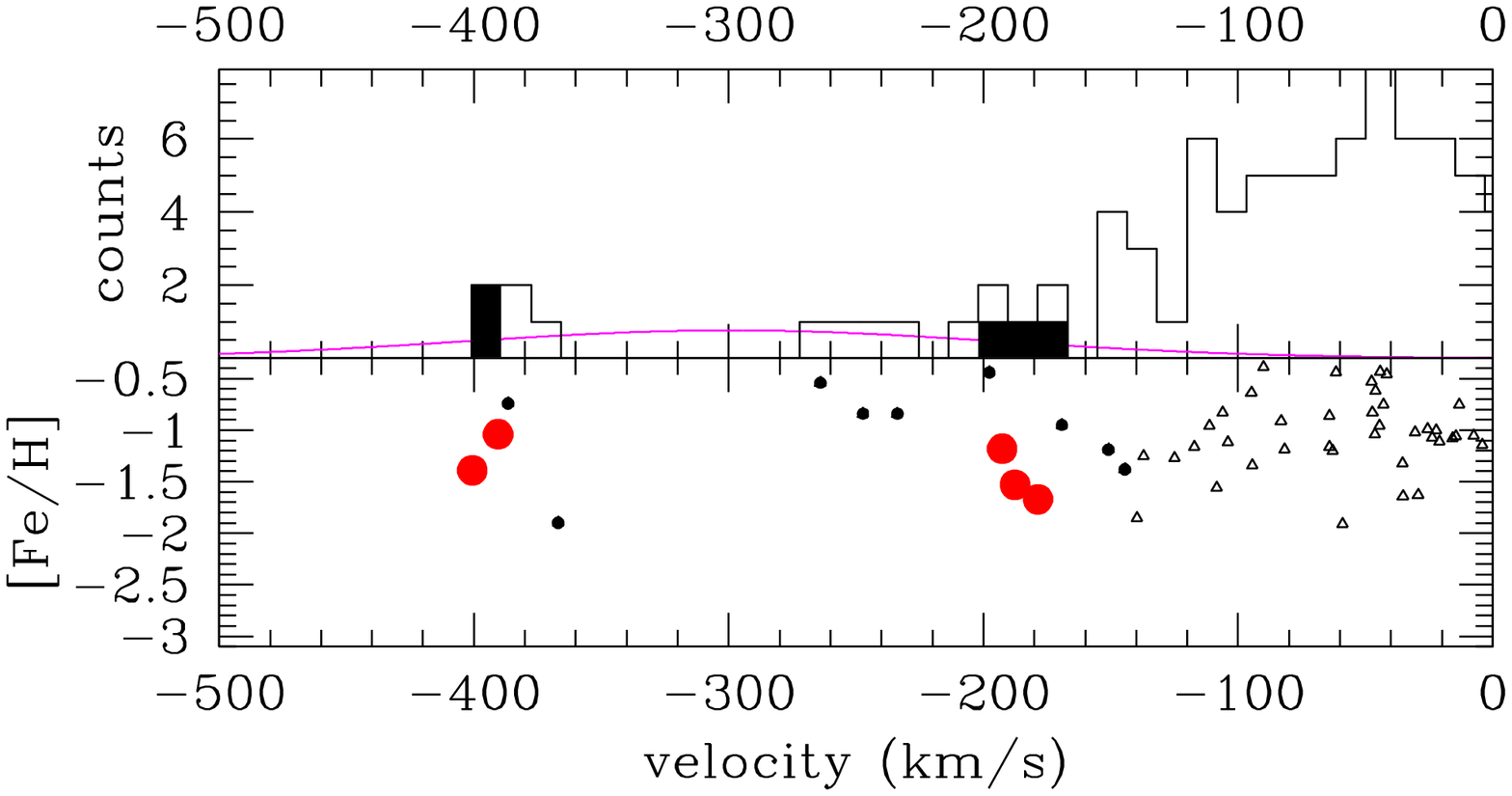}
\caption{
The same as for Fig.~4 for a field further along the \streamd\ structure (F37).
Symbols are again highlighted which have photometric metallicities
consistent with the [Fe/H] distribution of \streamd\ in Ibata et al.\ (2007): $-1.7 < [Fe/H] < -0.7$.
}
\end{center}
\end{figure*}

\begin{figure*}
\label{streamb}
\begin{center}
\includegraphics[angle=0,width=0.49\hsize]{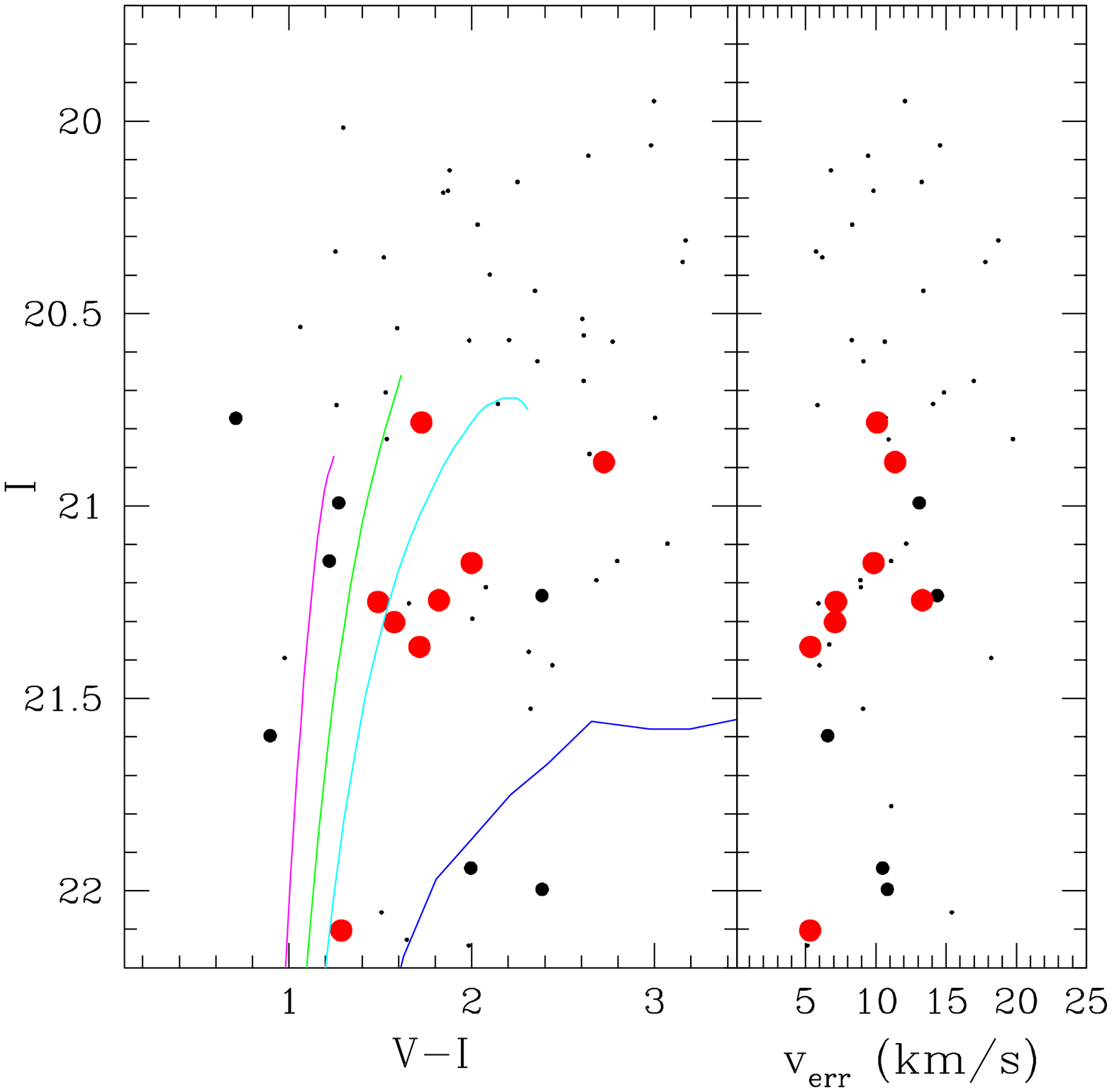}
\includegraphics[angle=0,width=0.49\hsize]{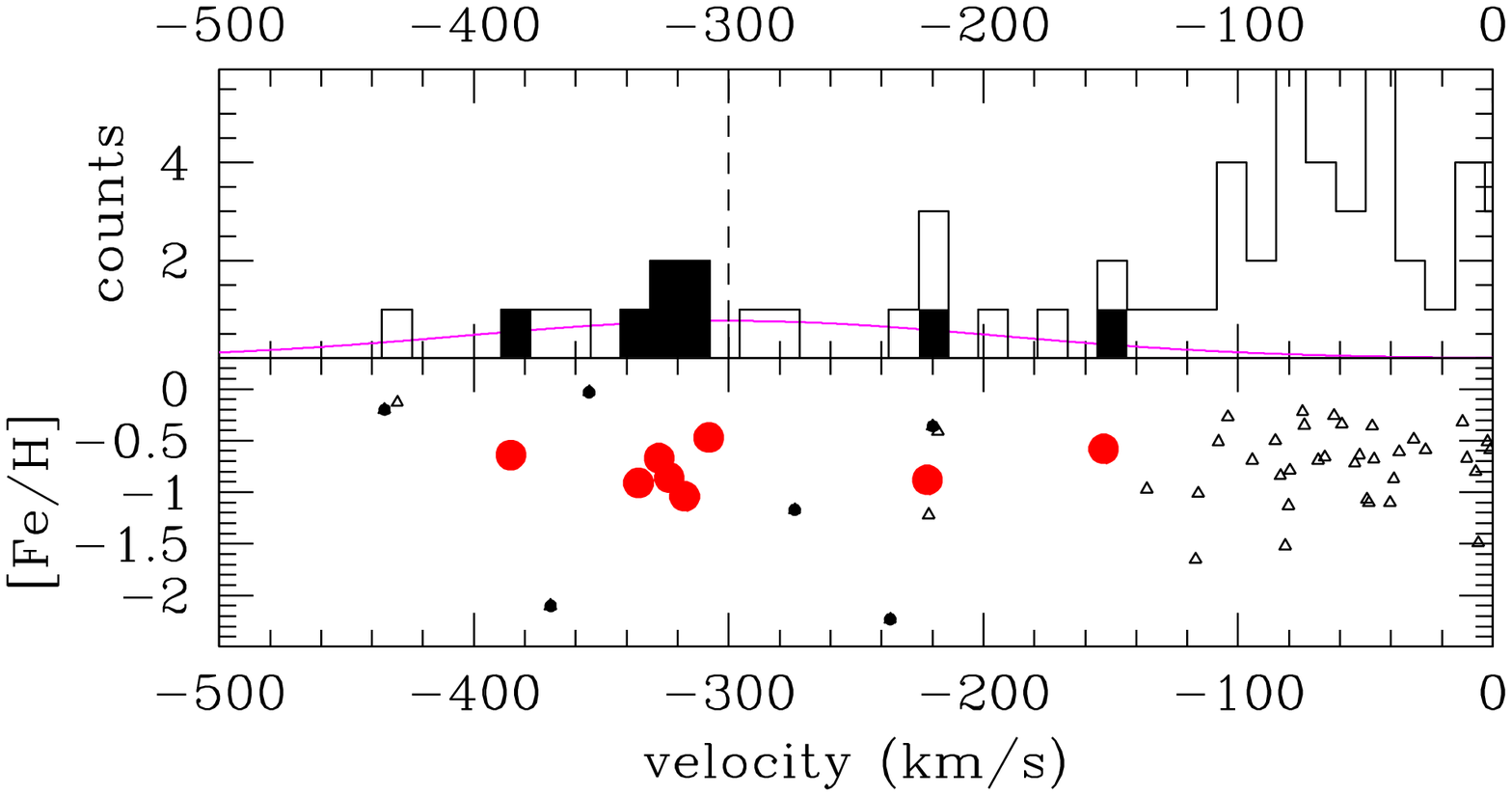}
\caption{
The same as for Fig.~5 for stars in the \streamb\ field, named `M6' from Gilbert et al.\ (2006) and Koch et al.\ (2008).
Symbols are again highlighted which have photometric metallicities
consistent with the [Fe/H] distribution of \streamb\ in Ibata et al.\ (2007).
}
\end{center}
\end{figure*}

\begin{figure*}
\label{streama}
\begin{center}
\includegraphics[angle=0,width=0.49\hsize]{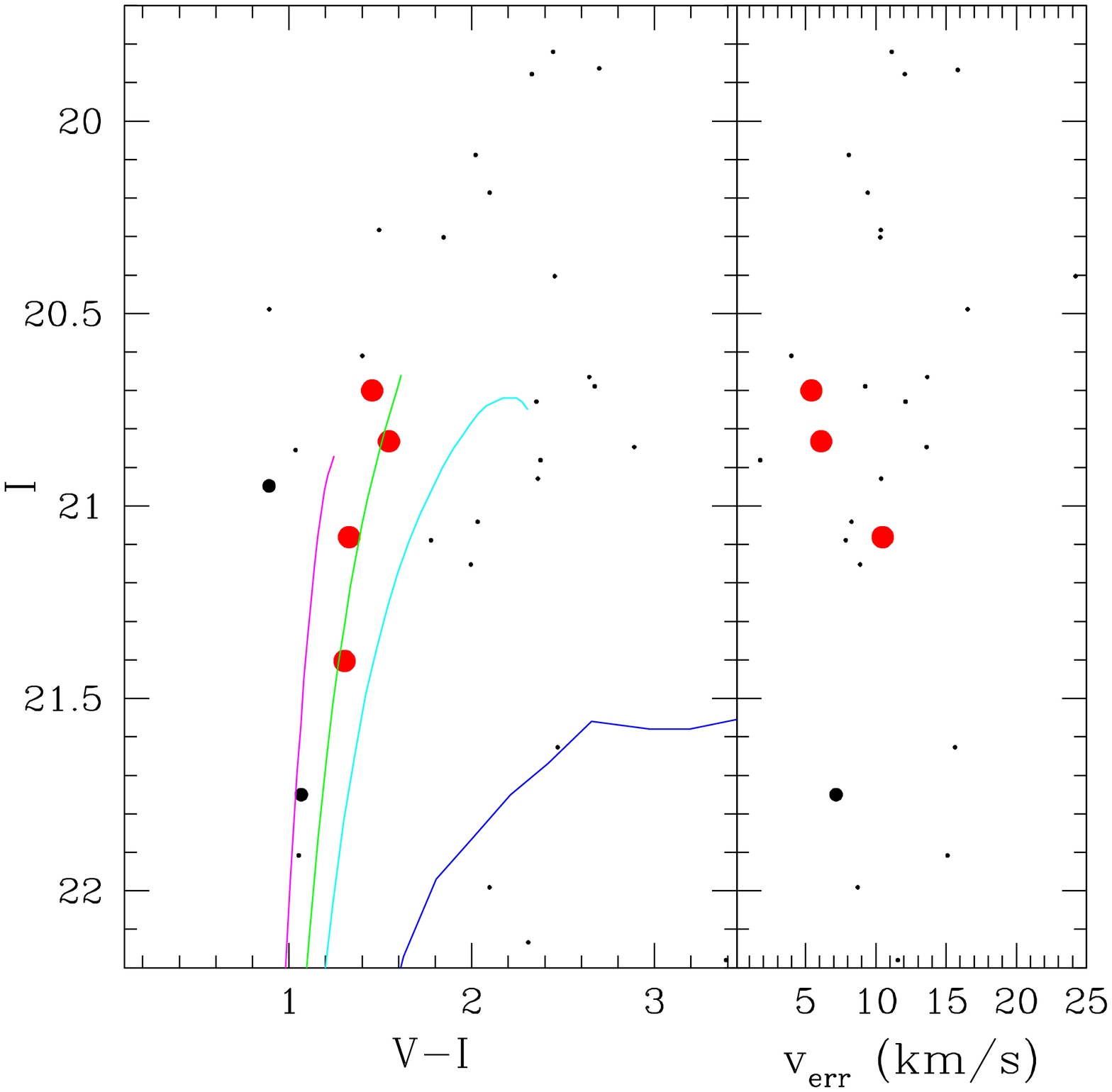}
\includegraphics[angle=0,width=0.49\hsize]{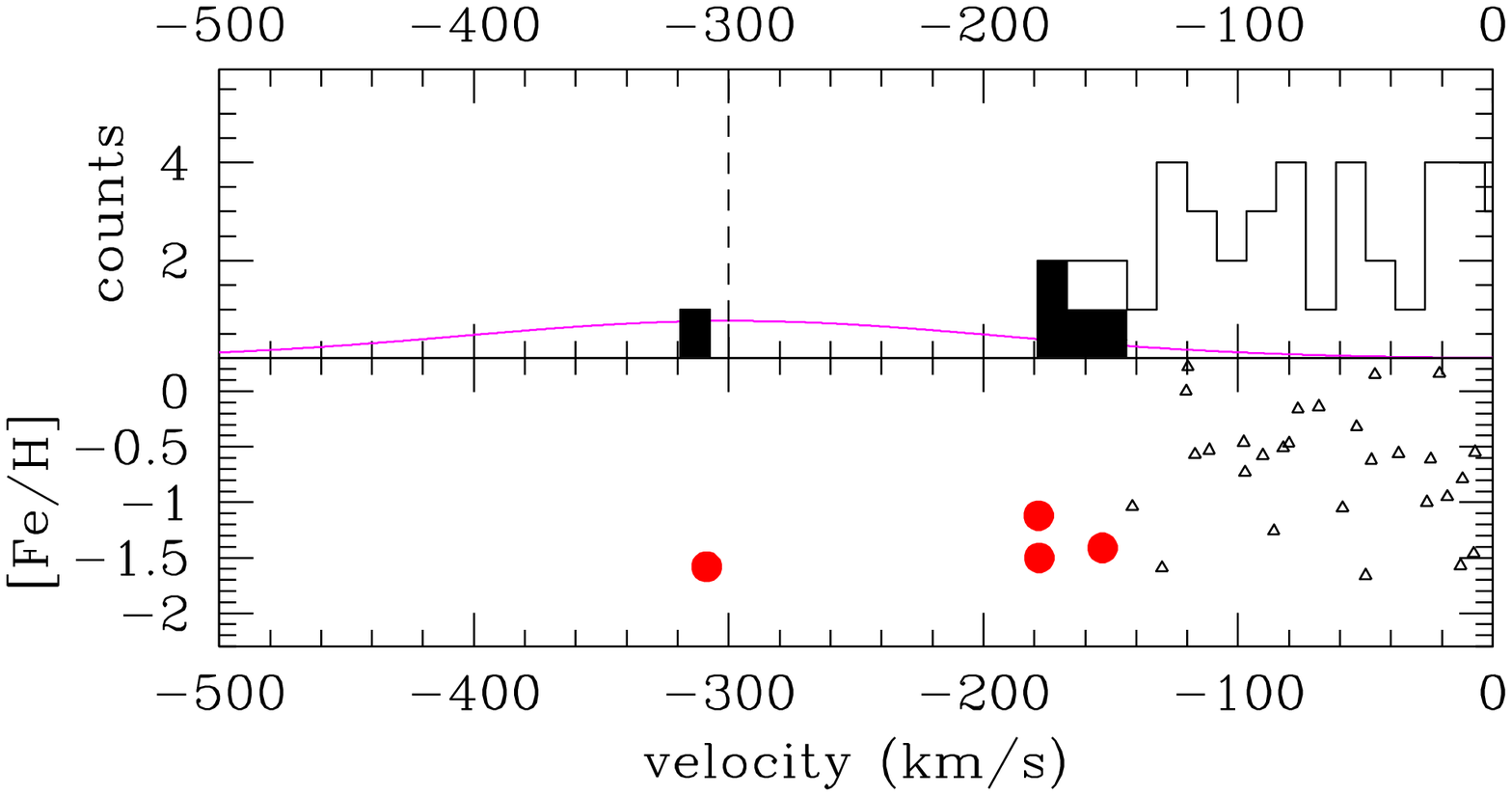}
\caption{
The same as for Fig.~5 for stars in the \streama\ field, named `M6' from Gilbert et al.\ (2006) and Koch et al.\ (2008).
Symbols are again highlighted which have photometric metallicities
consistent with the [Fe/H] distribution of \streama\ in Ibata et al.\ (2007).
}
\end{center}
\end{figure*}

\subsection{Velocity accuracy: repeat measurements of stars in fields 
F25 \& F26}

While CaT fitting errors suggest relatively small velocity errors,
an independent check can be made on the 56 radial velocities of stars
lying in spectroscopic masks of both fields F25 and F26.
Velocity differences are shown in Fig.~8 
 highlighting those corresponding to 
\streamc\ (3 stars), \streamd\ (2 stars), \ec4\ (3 stars), background halo (2 stars), and Milky Way foreground (46 stars).
A systematic shift from night\,1 to night\,2 of 3~km/s was found over all velocities, and this has been removed as a constant. Agreement between observing nights for these stars is then generally 
found within the 1$\sigma$ errors of the radial velocity measurements, 
suggesting that no significant skew from mask misalignments or systematic errors are present from
instrumental setup night to night.
The dispersion in velocity differences is $\sim$6~km/s for both the M31 sample and the Milky Way sample taken separately, which is comparable to the typical velocity measurement error of an individual star.
For these 56 stars, we have taken as the radial velocity the error-weighted average of the  measurement from the two nights.

\begin{figure*}
\label{streamc}
\begin{center}
\includegraphics[angle=0,width=0.43\hsize]{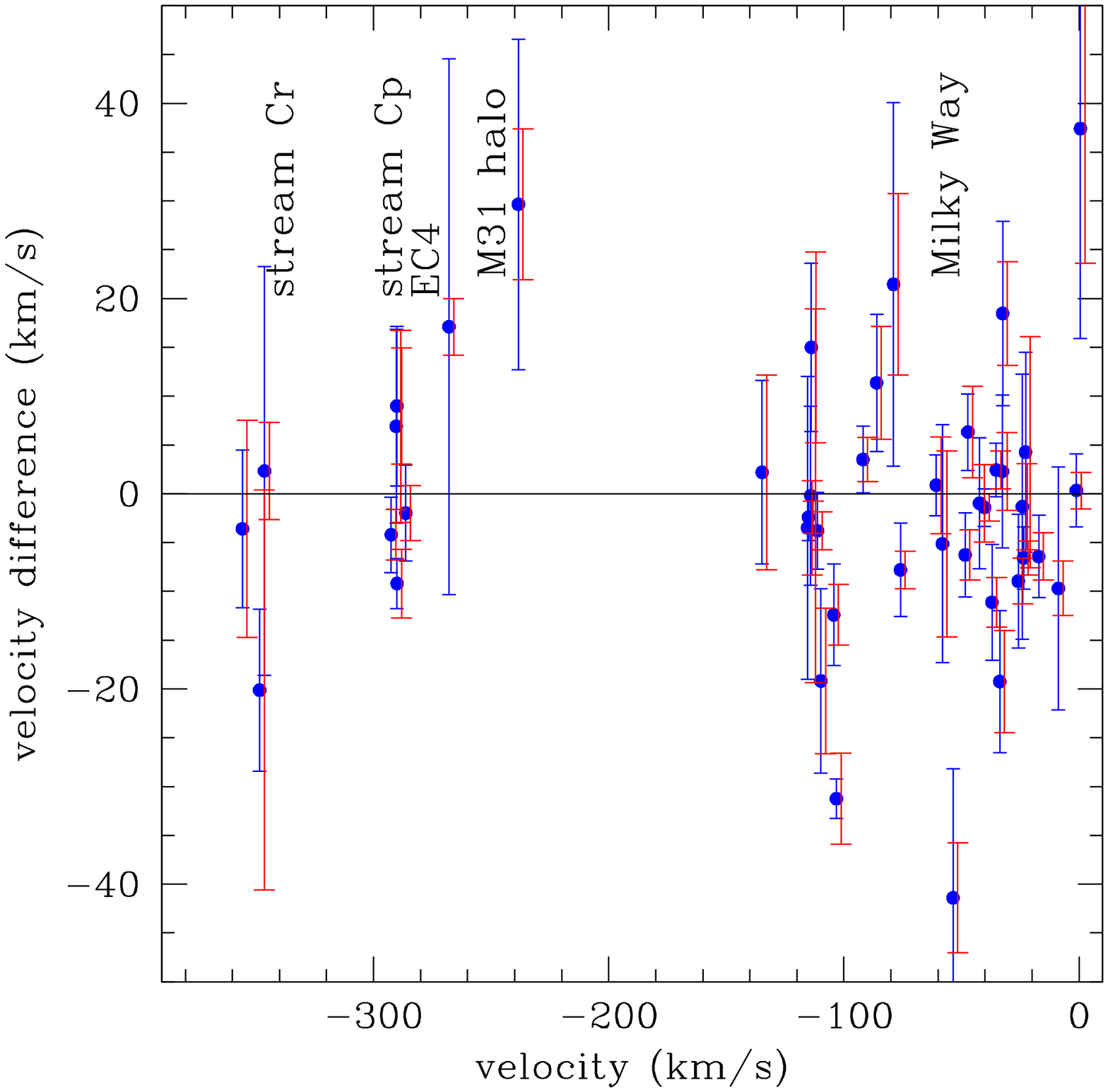}
\caption{ 
Velocity differences of stars lying in both fields F25 and F26.
Stars identified to velocity regions likely associated to 
\streamcr, \streamcp, \ec4, background M31 halo, and Milky Way foreground 
are identified.
Stars are shown at their velocities from mask F25, with offset error-bars
from mask F26.
}
\end{center}
\end{figure*}


%
\section{Results}

\subsection{Global kinematics and metallicities of the fields}

Figures~2--7 (left panels) show the
CFHT-megacam colour-magnitude diagrams (CMD)  and radial velocity uncertainties
of the observed stars in the \streamc, `D', `B', and `A' fields, 
while the right panels show velocity histograms and photometrically derived metallicities,
[Fe/H] = $\log(Z/Z_\odot)$, 
are computed for the stars by interpolating between 10\,Gyr old 
Padova isochrones (Girardi et al.\ 2004).
The fixed age adopted for metallicity comparison to the fiducials 
will of course introduce a systematic uncertainty if they aren't all old. 
Given that younger populations have been detected in the halo of M31, and that these streams could represent progenitors with a range of properties, we provide an estimate of the  age variations
on the [Fe/H] determinations. If 5\,Gyr old isochrones are used, the apparent metallicity would shift by 
0.2~dex more metal-rich. 
The average distance modulus of M31 is adopted, 
24.47, 785~kpc (McConnachie et al.\ 2005).
Stars which are unlikely to be contaminated by foreground
Milky Way ($v_{hel}<-150$~km/s) are highlighted.
While spectroscopic metallicities are quoted in the Tables, we do not use them for analysis here as the errors on individual star measurements are so large as to broaden the typical [Fe/H] distribution by a factor of three for a stream kinematic structure.
While the photometric [Fe/H] determinations are highly model dependent, the distributions for a given structure are likely far more reliable than those measured from the relatively low-S/N spectroscopy.


\subsection{Stream `Cr'}

\streamc\ is the dominant stellar component at the position of our 
spectroscopic masks F25/F26, exceeding the halo stars as well as stars likely to be foreground MW contaminants (in the adopted velocity range of the M31 halo: $<$-150~km/s)
by a factor of $\sim$3$\times$ on average. 
In Fig.~2, the stars likely belonging to \streamc, and \ec4\ are highlighted (where EC4 member stars were preferentially inserted in the spectroscopic masks)
allowing metallicity comparison to the fiducial globular cluster RGBs.
To differentiate EC4 stars from the halo field,
the velocities are also plotted against their radius from the \ec4\ center.
Figure~2 
shows that once the cluster, \ec4, stars are removed, 
a strong metal rich peak of stars at $\sim-350$~km/s dominates the stars 
kinematically identified to exclude the Milky Way.
The stars in this kinematic structure have  average photometrically derived metallicity, $[Fe/H]$=-0.74$\pm$0.19. %
As the metallicity of this kinematic substructure
is very similar to that measured for the total 
\streamc\ ([Fe/H]=-0.6) by Ibata et al.\ (2007), this is an excellent candidate for a kinematic detection
of \streamc.
Taking the clump of metal rich stars $\pm2\sigma$ from the peak, 
we find nine stars with an average
$<v_r>=-349.5^{+1.8}_{-1.8}$~km/s, $\sigma_{v_r}=5.1^{+2.5}_{-2.5}$~km/s, where the individual
velocity errors are taken into account in a maximum-likelihood sense. This procedure is described exactly in Martin et al.\ (2007), although no iterative clipping is done in this case as the contamination levels from foreground Milky Way stars is much smaller. Briefly,
using only the candidate \streamcr\ stars,  a maximum-likelihood algorithm that
explores a coarse grid of the $(v_r,\sigma)$ space and searches for
the couple of parameters that maximizes the $ML$ function\,\footnote{There is an error in this expression from Martin et al.\ (2007) which is corrected here.} defined as:

\begin{equation}
ML(v_r,\sigma)=\sum_{i=1}^{N}\log\Big(\frac{1}{\sigma_{\mathrm{tot}}}
\exp\Big[-\frac{1}{2}\big(\frac{v_r-v_{r,i}}{\sigma_{\mathrm{tot}}}\big)^2\Big]\Big)
\end{equation}

\noindent with $N$ the number of stars in the sample, $v_{r,i}$ the radial
velocity measured for the $i^\mathrm{th}$
star, $v_{err,i}$ the corresponding uncertainty and $\sigma_{\mathrm{tot}}=\sqrt{\sigma^2+v_{err,i}^2}$. Using this
definition of $\sigma$ allows to disentangle the intrinsic velocity dispersion 
and the
contribution of the measurement uncertainties to these likelihood distributions\,\footnote{An alternative way to determine
$v_r$ and $\sigma$ is to use $\sigma=\sigma'_{\mathrm{tot}}$ in equation (1) to measure the observed dispersion and then
correct from the mean velocity uncertainty, $\overline{v_{err}}$, such as
$\sigma'_{\mathrm{tot}}=\sqrt{\sigma^2+\overline{v_{err,i}}^2}$. Parameters obtained in this way are similar to those given in the text.}.
These distribution functions are shown in Figure~9.

\begin{figure*}
\begin{center}
\includegraphics[angle=270,width=0.605\hsize]{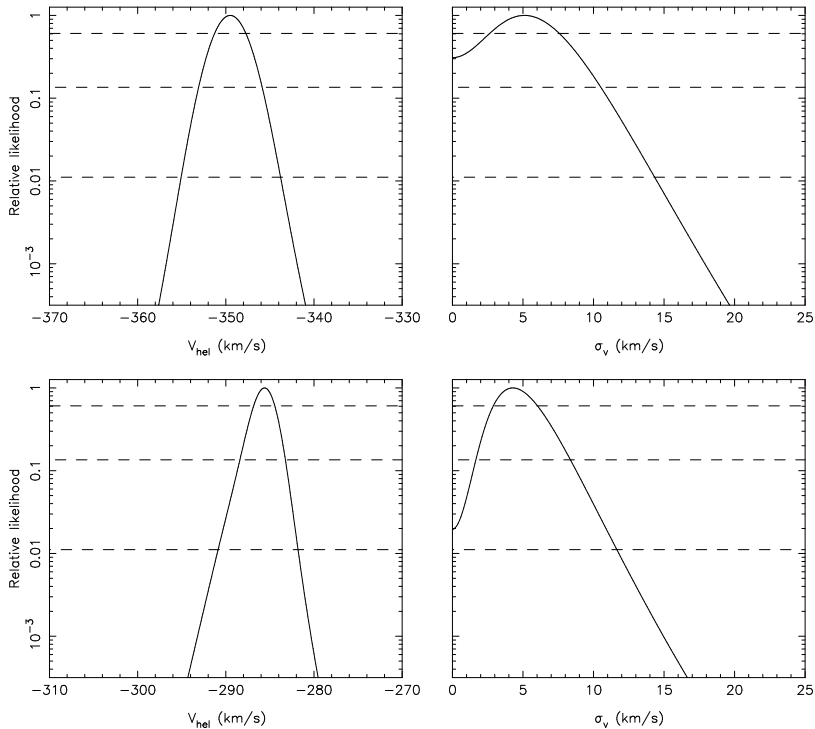}
\end{center}
\caption{The kinematics of \streamc.
{\bf Upper panels:} 
The relative likelihood distribution (taking into account the measurement errors) of $<v_r>$ and 
$\sigma_{v_r}$  for velocities of \streamcr\ stars
when marginalizing with respect to the other parameter. The thin  
dashed lines correspond, from top to bottom, to the parameter range  
that contains 68.3\%, 95.4\% and 99.73\% of the probability distribution
(1, 2 and 3$\times$ $\sigma$ uncertainties),
revealing that 
\streamcr\ has a none-zero velocity dispersion at the $1\sigma$ level, and a remarkably small
$\sigma_{v_r}$.
The peak values and $1\sigma$ range are $<v_r>=-349.5^{+1.8}_{-1.8}$ and 
$\sigma_{v_r}=5.1^{+2.5}_{-2.5}$.
{\bf Lower panels:} 
The same for \streamcp\ (with 6 candidate members), showing a none-zero velocity dispersion at the $>3\sigma$ level.
The peak values and $1\sigma$ range are $<v_r>=-285.6^{+1.2}_{-1.2}$ and 
$\sigma_{v_r}=4.3^{+1.7}_{-1.4}$.
Note that both streams have maximum likelihood distributions close to symmetric about the solution in projection.  
}
\label{MLplots}
\end{figure*}

The velocity is close to the systemic velocity of M31, -300~km/s, which might be expected if the stream were truly close to tangential as it appears to be in the imaging (most of its velocity being orthogonal to
our measured heliocentric component).
We cannot estimate a reliable mass for the progenitor from this single spectroscopic measurement, as the stars we have detected lie off-center
from the \streamc\ peak, and as yet we have not measured the full extent of the stream.
Nonetheless, we can at least place a constraint on the mass from the measured
velocity dispersion in this field (\S~4). 

By taking the average halo profile of Ibata et al.\ (2007) at this projected
radius, $\sim$9 true halo stars are expected in fields F25/F26 (all
stream structures removed from consideration),
assuming all possible candidate RGBs have been observed in the
two overlapping DEIMOS pointings, which they have.
We find 26 halo stars are detected in the velocity region 
v$_{\rm hel}<-150$~km/s excluding the MW, and after removing the high confidence 
EC4 stars (lying within 2 core radii)
from consideration as they were added
selectively to the mask in addition to the randomly selected halo stars.
If 9 stars are associated to \streamcr\ and as we shall see another 7 stars
are associated to \streamcp, there are 10 candidate halo stars found
in the sample, in good agreement with the average prediction. 
Integrating the windowed $\sigma_v\sim125$~km/s Gaussian of the halo, 
we find 3\% chance that a star lies in one of the three 10~km/s velocity 
bins encompassing this kinematic structure.
At least one, but unlikely more than two of the candidate \streamcr\ stars will
be unrelated halo.
Therefore neither the metallicity or velocity dispersion is likely to 
be heavily biased by unrelated halo stars.
We note that $<<1$\% Galactic contamination is expected at $\sim-350$~km/s,
from our own characterization of the MW population in our spectroscopic fields, 
from the Gilbert et al.\ (2006) analysis of MW dwarfs in their M31 spectroscopy, and from 
the Besan{\c  c}on Galactic populations model (as described in Ibata et al.\ 2005, 2007 and Chapman et al.\ 2006).

\subsection{Stream `Cp'}

In analyzing the velocity distribution of halo stars in field F25/26,  we move on from the strong kinematic peak of stars at
-349~km/s which we have identified with \streamcr, and notice in figure~2 a kinematic association of stars at v$_{\rm hel}\sim$-286~km/s showing a remarkably small dispersion. 
Figure~2 also plots the radial distance of the stars in the field from the cluster \ec4, where it is apparent that this spike merges with likely \ec4\ member stars. There are five stars lying within one core radius of \ec4\ ($<$30\,pc), two borderline members between 2--3 core radii, and 6 candidate stream stars at such large distance that they are unlikely to be directly associated to \ec4.

We note that one of these stars is likely either a blue supergiant star at M31 distance or is a Milky Way contaminant, although the equivalent width of the Na{\sc I} doublet is remarkably small for a Milky Way dwarf (e.g., Guhathakurta et al.\ 2006, Martin et al.\ 2007), with the caveat that this example is very blue in colour where it is not clear Na{\sc I} is a good discriminant (e.g., Koch et al.\ 2008). 
Since no photometric metallicity can be derived for this star, we will not consider it further in our analysis (the star is \streamcp, 00:58:24.32+38:04:29.9).

A tight range in [Fe/H]=-1.26$\pm$0.16 is observed in the six unambiguous \streamcp\ stars, 
very close to what we derive from the Ibata et al.\ (2007) foreground subtracted photometry dataset for the offset  \streamcp\ region, [Fe/H]=-1.1.
We propose that in addition to \streamcr, we have also kinematically detected this lower contrast \streamcp\ in our fields F25/F26.
It is a matter of debate how the two borderline \ec4\ stars are treated; starting from the \streamcp\ standpoint we find no valid reason to reject them from the stream sample, as they lie well within the velocity window defined by the stars at much larger radius (see Fig.~2), although we quote our results with and without them.
Again, the systemic velocity close to that of M31 (v$_{hel}$=-285.6$\pm1.2$~km/s with 6 members,
v$_{hel}$=-285.0$_{-1.8}^{+1.7}$~km/s with 5 members) would be consistent with the expected properties of a tangential stream.
Using the maximum likelihood technique we calculate a true velocity dispersion of 
$\sigma_{v,corr}=4.3_{-1.4}^{+1.7}$~km/s with 6 members ($\sigma_{v,corr}=5.1_{-2.5}^{+2.5}$~km/s with 5 members).
The likelihood distribution functions are shown in Fig.~9.

\begin{figure*}
\begin{center}
\includegraphics[angle=0,width=0.505\hsize]{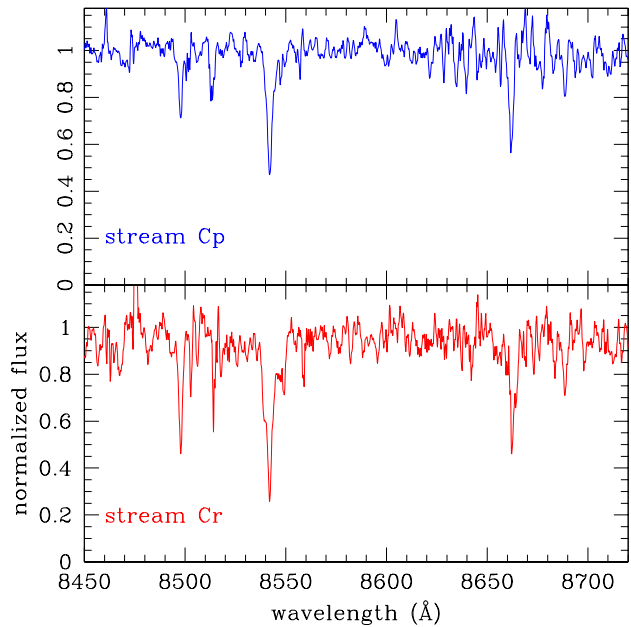}
\end{center}
\caption{The stacked spectra of \streamcr\ and \streamcp, weighting by the inverse variance in the continua, emphasizing the clear spectroscopic difference in [Fe/H] (-0.8 versus -1.4) between the two kinematic peaks, in good agreement with the photometric [Fe/H].
}
\label{spec}
\end{figure*}

The stacked spectra of \streamcr\ and \streamcp\  are shown in Fig.~10, emphasizing the clear difference in spectroscopically derived [Fe/H] (-0.8 versus -1.4) between the two kinematic peaks, in good agreement with the photometric [Fe/H] quoted in Table~1 (despite the fact that individual spectra are low signal-to-noise and show a large spread in [Fe/H]).    
With confidence that we have truly detected two different, superposed stream components through their offset kinematics and metallicities, we revisit the \streamc\ region from Ibata et al.\ (2007). 
In Fig.~11 we show a zoomed-in region around \streamc, divided in [Fe/H] into two non-overlapping ranges (0.0 to -0.7, and -0.7 to -1.7). This division makes it clear that two physically  separate structures are present, and we can attempt to separate the luminosities of the two components (see \S~4).

\begin{figure*}
\begin{center}
\includegraphics[angle=270,width=0.495\hsize]{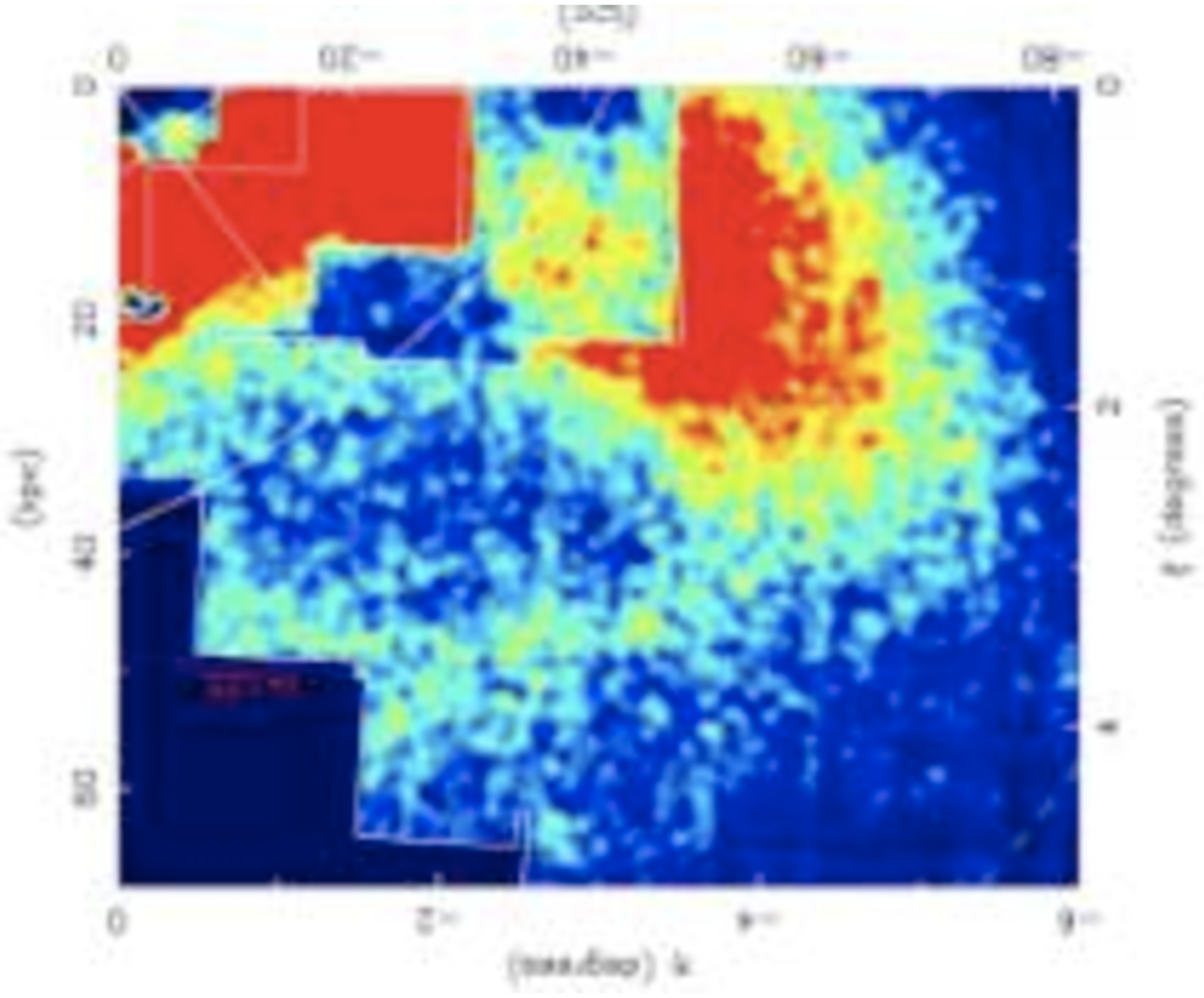}
\includegraphics[angle=270,width=0.495\hsize]{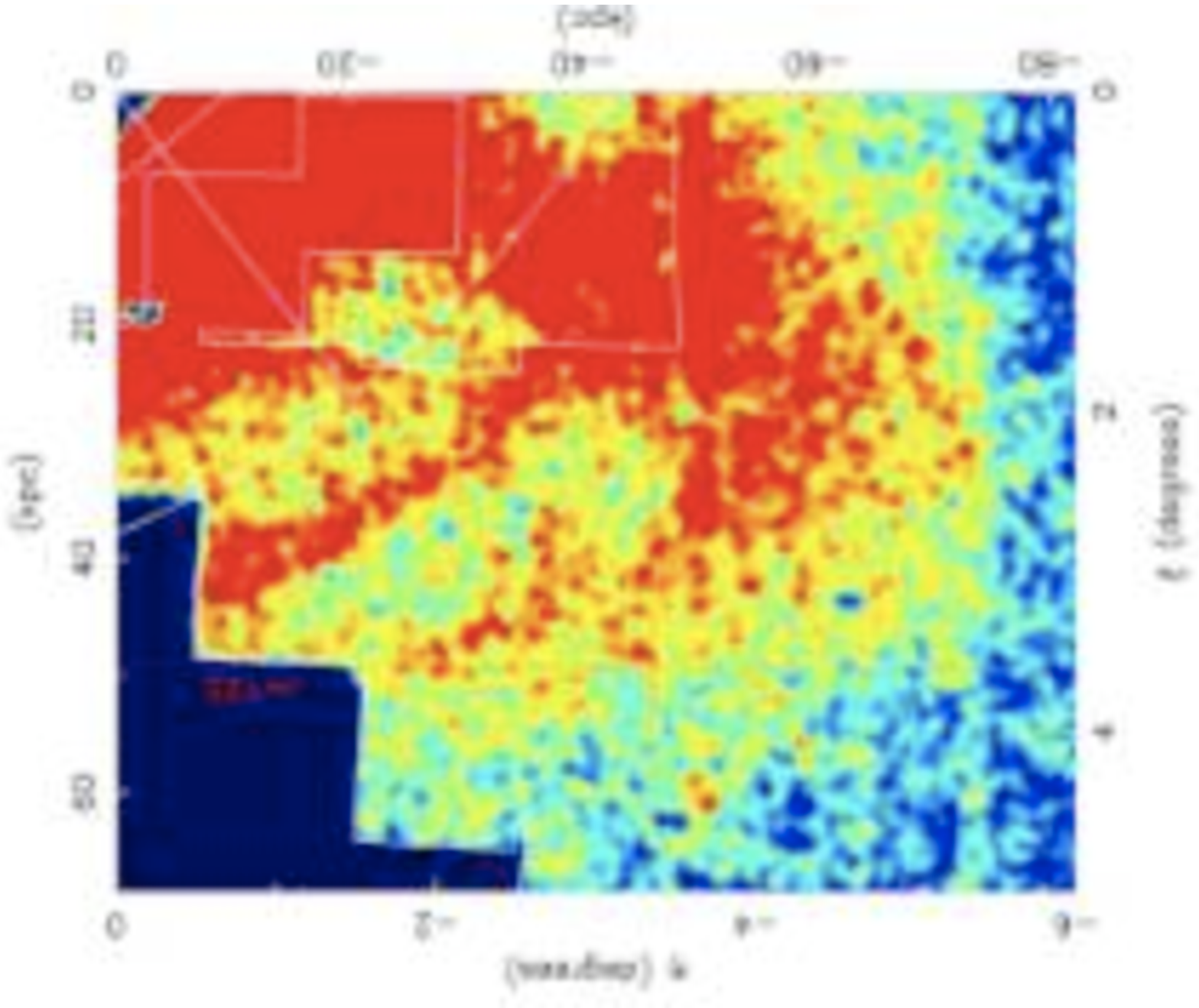}
\end{center}
\caption{The \streamc\ region shown zoomed-in (from Fig.~20 of Ibata et al.\ 2007, which details the definitions of cutout regions etc.)
with slices in [Fe/H]  ($-0.7<[Fe/H]<0.0$ -- left panel; $-1.7<[Fe/H]<-0.7$ -- right panel)
allowing the spatially offset
\streamcr\ (left panel) and \streamcp\ (right panel) components to clearly be seen. Our kinematic separation of these two components appears to be reflected in physically distinct (but somewhat overlapping) regions. The comparison also highlight the much more metal poor \streamd\ in the right panel.
}
\label{zoom}
\end{figure*}

With conservatively only 5  member stars (the 6th not obviously being an RGB star), and a metallicity range well within that expected for     
M31's overall stellar halo ([Fe/H$\sim-1.4\pm0.2$ from Chapman et al.\ 2006;  Fe/H$\sim$-1.2 along the minor axis in Kalirai et al.\ 2006 or Fe/H$\sim$-1.5 over the same minor axis fields in Koch et al.\ 2008),
we should first consider how likely this kinematic peak is to be distinct
from the smooth halo component.
It is also of interest  to demonstrate that these stars are not likely 
to be far-flung members of \ec4.

These five stars at large radii from \ec4 (0.7 to 10.3~arcmin, or 150\,pc to 2220\,pc), chosen by chance in the spectroscopic masks are unlikely to be bound members of \ec4\ itself.
Firstly, these stars would represent 5-74 core radii of \ec4 (30~pc).
It is implausible that
stars this far from the center belong to the cluster unless
it is strongly disrupted, or is in fact only the cold core component
of a more diffuse dwarf galaxy with an outer second component of stars.
\ec4 does not show any obvious signs of disruption, but given the 
faintness of \ec4, it is difficult to tell from the HST/ACS image
in \citet{mackey06} whether there are subtle signatures of disruption.
It can also be seen in Fig~2 
that the metallicities of these stars appear to be marginally richer than \ec4, although photometric errors from EC4 star crowding
could easily account for these differences.

It is more difficult to differentiate these stars from M31 halo stars than it is for \streamcr\ stars, since the
peak of the halo velocity distribution lies at $\sim$-300~km/s,
although the very broad velocity dispersion $\sim$125~km/s at this
projected radius \citep{chapman06} makes it less likely to find such a strong
spike of stars at -286~km/s.
We estimate the chance association specifically as follows.
%
%
We assume we know nothing about \ec4\ and that we have an ensemble of
stars which are members of the M31 halo. For this purpose we assume
conservatively that a halo star is
any star consistent with the halo CMD, and to coarsely
remove Galactic contaminants, lying within
$-550$~km/s $< v_r < -150$~km/s, or roughly $\pm2\sigma$ with a window clip
appropriate to the Galactic contaminant distribution in Fig.~2. 
Removing the stars specifically targeted to lie in \ec4, there are 26
stars which satisfy these  criteria.
There is a 3.2\% chance that any given halo star will lie in the $10$~km/s
window centered on the five unambiguous \streamcp\ stars systemic velocity (-285~km/s). 
This is a conservative assessment since the chance would 
be much lower for a window offset from the $\sim$-300~km/s peak of the
M31 halo distribution.
However, the chance that 5 stars out of 26 lie in this window is extremely
small (irrespective of how the window is defined, the probability is consistently below $<10^4$), given the broad halo $\sigma_v$.
This estimate is even more conservative as there is in fact a
substantial contribution from the \streamcr\ stars
in this halo field -- a better estimate of the total underlying halo stars would be $\sim16$.
We can conclude that these 5 stars represent a rare kinematic spike in a smooth
halo distribution.

\subsection{The continuations of \streamc}

As described in \S~2, upon detecting cold kinematic peaks plausibly attributed to \streamc, we observed
an additional Keck/DEIMOS field (F36) further North along this structure.
The properties of this field are shown in Figure~3.
While no prominent peaks are found in the velocity distribution of this northern field, a concentration of  five metal-rich ([Fe/H]$\sim$-0.7) stars is observed at an average velocity of -350~km/s, possibly attributable to the same \streamcr\ structure, and thereby showing no obvious velocity gradient. 

In the same field (F36) we also search for the more metal-poor \streamcp. We find a clump of four similarly metal-poor stars, $<$[Fe/H]$>$=-1.2, at an average velocity of -246~km/s, offset by $\sim+40$~km/s from the \streamcp\ and \ec4\ peak in fields F25/26.

 Orbit models of these streams will be presented in a future paper, while Fardal et al.\ (2008) discuss how these stream-like structures could conceivably be related to the Giant Southern Stream.

\subsection{The extended star cluster, EC4}

Figure~2 shows that stars targeted in the cluster, 
\ec4, have clearly been kinematically identified with a distribution of 
velocities
centered at v$_{\rm hel}$=-285~km/s. 
A detailed study of EC4 (Collins et al.\ in preparation) suggests a small ($\sim3$\,km/s) resolved
velocity dispersion, marginally consistent with a dark matter dominated system.
The \ec4\ stars have [Fe/H]$_{phot}$=-1.4$\pm$0.1, although a spectroscopic estimate of the [Fe/H]=-1.6$\pm$0.15\, \footnote{Mackey et al.\ (2006) find  [Fe/H]$_{phot}$=-1.84 for \ec4\ from HST photometry and simultaneous fitting of the RGB and Horizontal Branch. This difference is explored in Collins et al.\ (in preparation).}.
As these stars were preselected to lie in \ec4, we assume that at least the 5 stars lying within 1 core radii of the \ec4\ center (and likely  the two stars at 2--3 core radii) can be removed from the surrounding M31 halo sample for our statistical analysis of the \streamc\ in previous sections. 
We have clearly identified the 
systemic velocity of \ec4\ as being compatible with the \streamcp\ 
kinematics. 

\subsection{Stream `D'}
Based on the example of \streamcr\ and \streamcp, we are motivated to search for a
narrow velocity peak in the case of \streamd. However Fig.~4 
showing the CMD, radial velocities and metallicities (as for
the previous streams) does not reveal any obvious
detection of stars in this stream.
We proceed by comparing the expected metallicity from Ibata et al.\ (2007), $-1.7 < [Fe/H] < -0.7$, with any stars in the ``halo" sample (culled from the velocity and EW(Na{\sc I}) cuts) which could be the \streamd.
We highlight all stars in the CMD which could conservatively be consistent with the
\streamd\  median [Fe/H]=-1.2. There are no obvious kinematic spikes within these
colour-selected stars, as shown in Fig.~4 (right panel). 
However, there is an isolated 
group of two stars within a 10~km/s bin at -405~km/s.
Two stars at $\sim$-400~km/s are somewhat unexpected (8\% chance) given the 
halo velocity dispersion at a projected radius of 35~kpc (Chapman et al.\ 2006).
There are no better candidates for the \streamd\  than this pair of
RGB stars, but we cannot confidently separate \streamd\  stars from
spheroidal halo stars.

In the DEIMOS field placed further along  \streamd\ (field F37),  an kinematic  association of 5  stars stands out again at $\sim$400~km/s (as with field F7 above), however only 3 have inferred [Fe/H] within a range consistent with the photometric properties/ If these stars represent the \streamd, there is also no measurable velocity gradient detected, as found for \streamcr.
In Table~1 we present the velocity dispersion measured from the 5 plausible \streamd\ stars combined from both fields ($\sigma_{v_r}$=4.2~km/s).
However, the low contrast of \streamd\ relative to the background M31 halo, together with our inability to distinguish with confidence a kinematic identification mean that these results are tentative.
Much larger numbers of spectroscopic measurements along \streamd\ are required in order to reliably detect a coherent structure in velocity. 

\subsection{Streams `A' \& `B'}
The other two streams lying perpendicular to the minor axis presented in Ibata et al.\ (2007), \streama\ and \streamb\ lie at 120~kpc and 80~kpc respectively.
Both of these streams have serendipitous spectroscopic pointings lying in their edge regions from
Gilbert  et al.\ (2006) and Koch et al.\ (2008), named fields M8 and M6 respectively in their nomenclature
(see Ibata et al.\ 2007 for placements of these spectroscopic pointings in the wider M31 halo map).
We present the CMDs, velocity histograms and metallicities for \streamb\ and \streama\ here for analysis (Figs.~6 \& 7).

In field M8 (\streama) there are only 4 stars with velocity measurements attributable to the M31 halo.
The clump of three stars at $-172$~km/s have an [Fe/H]=-1.3 on average, very similar to the 
statistical measurement of the [Fe/H]$\sim-1.3$ in \streama\ (Ibata et al.\ 2007).
We further suggest, along Ibata et al.\ (2007) and Koch et al.\ (2008), that these 3 kinematically identified {\it halo} stars are more likely to be associated to the stellar halo of M33 (the Triangulum galaxy), given the M31 halo velocity dispersion (Chapman et al.\ 2006) shown in Figs.~6 \& 7.
It is therefore worth considering that this \streama\ structure might actually be disrupted remains of a satellite in M33's halo.
The dispersion of these three stars, 14.5~km/s, cannot easily be deconvolved  for measurement errors using the maximum likelihood approach. Instead, since the errors are similar for all three stars, we
write as in footnote(1) $\sigma_{v,corr} = \sqrt(\sigma_v^2 - \sigma_{instr}^2)$ $= \sqrt(14.5^2 - 7.2^2) = 12.5$~km/s.

We carry out the same procedure as with the other streams, identifying candidate stars in the CMD which are consistent with the average metallicity found in Ibata et al.\ (2007) for the streams in question.
In field M6 (\streamb), a kinematic peak of stars lying at $\sim$-330~km/s with
 $<$[Fe/H]$>\sim-1.0$ represents a reasonable candidate for  this stream. Notably, this metallicity is very close to that estimated for \streamb\ in Ibata et al.\ (2007), and further, both the metallicity and the velocity distribution in this field depart significantly from the average found in Koch et al.\ (2008) for the outer halo. 
 Again following the logic of our discussions in the streams `C' and `D', the RGB overdensity in this field attributed to the stream should statistically result in the bulk of stars kinematically detected lying in the stream structure. Removing these five relatively metal-rich stars from the halo sample leaves six more metal-poor stars which would represent the surrounding M31 halo at this radius.
As with the \streama\ case above, we simply estimate an intrinsic dispersion directly from these five candidate \streamb\ stars, using their average measurement error, of $\sigma_{v_r}=6.9$km/s.

\section{The streams in the context of the dwarf spheroidal $M/L$ to $L$ relation}


It is of interest to ask how the properties of these streams would compare to other M31 satellites if treated as dwarf remnants. 
As described in Ibata et al.\ (2007) the light in polygonal regions surrounding each stream was integrated, with background corrections applied.
Here we go further to emphasize the metal-poor region of \streamc\ defined as the $V$-band light with
$-2.0<$[Fe/H]$<-1.0$, and further assuming (through the luminosity ratios of the offset portions of the metal-poor and metal-rich components, described above)
that 1/10 of this metal-poor light belongs to 
\streamcp\ while the other 9/10 belongs to \streamcr.
The resulting luminosities are summarized in Table~1.
There is of course an uncertainty in the M$_V$ estimates, since the streams are terminated by the edge of our M31 halo imaging to the North. 
We will therefore simply assume here that the M$_V$ estimates are lower limits.


\begin{figure*}
\begin{center}
\includegraphics[angle=0,width=0.595\hsize]{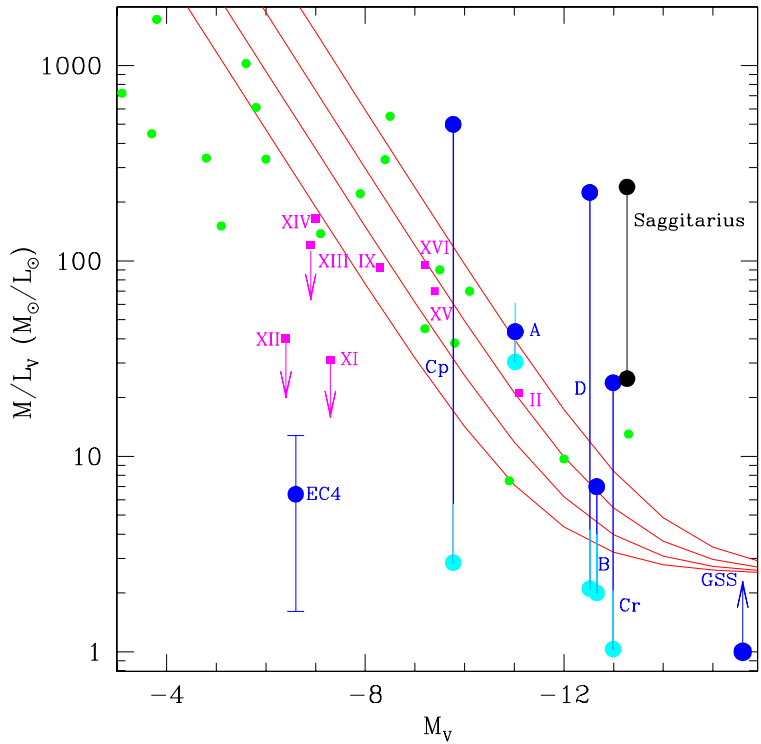}
\caption{
Comparison of the mass-to-light ratios (M/L) and
light, M$_V$ for the streams in M31, the Sagittarius stream in the Milky Way (Majewski  et al.\ 2003), and compared to the faint dwarf galaxies from
both the MW (Simon \& Geha 2007; Martin et al.\ 2007) and M31 with
central velocity dispersion estimates. 
The procedure for measuring the M/L is described in the text.
Stream M/L values are shown as bars connecting the structural mass estimate to the $\sigma_v$ 
 estimate. A factor of two uncertainty is shown for the $\sigma_v$ mass since we are likely observing the streams at especially cold  points between turning, and the true progenitor mass is likely 2--4 times larger.
\andxi, \andxii, and \andxiii\ (Collins et al.\ in preparation) are all shown as upper limits, since
their velocity dispersions are all unresolved by their measurements --
the 1$\sigma$ likelihood contour is used to set a tentative limit.
And{\sc XV} \& And{\sc XVI} measurements come from Letarte et al.\ (2008).
The solid lines are curves of constant dark matter halo mass (1, 2, 4, 8$\times$ 10$^7$ M$_\odot$ from bottom to top), assuming a stellar mass-to-light ratio of 2.5 M$_\odot$/L$_\odot$.
}
\end{center}
\label{ml}
\end{figure*}

The width of the streams can provide constraints on the mass of the progenitor, since debris from more massive 
satellites is likely to produce wider debris streams that spreads more rapidly 
along the orbit with time. 
We measure the widths of all our streams by taking the minor axis profile integrated over the full extent where the streams are detected. The streams are not always exactly orthogonal to the minor axis, however they are close enough (and in any case not always well defined) that any broadening introduced by this simple procedure should be minimal. We then fit Gaussians to the profile, subtracting the local background from regions on either side of each stream. We quote the FWHM in Table~1.

Johnston et al. (2001) present simple 
analytic scalings for the width and length of debris streams, with the main assumption that the progenitor is supported by random motions.  
In this case, the 
measured fraction, $s\equiv w/R$, of the width to the radial distance to the stream is related to the mass $m$ of the satellite through the relation 
$$s = (G m / [v_{circ}^2 R_{peri}])^{1/3}.$$

Font et al.\ (2006) estimate the progenitor mass of the Giant Southern Stream (GSS)  in M31 using these measurements ($w,R$) and an orbit model suggesting 
$R_{peri}=3-4.5~kpc$, finding $1.0-1.6\times10^8$~M$_\odot$. With an updated luminosity for the GSS from the wider survey of Ibata et al.\ (2007), $1.5\times10^8$~M$_\odot$, the M/L $\sim$ 1~M$_\odot$/L$_\odot$.
We note however that if the models of the GSS progenitor as rotationally supported (Fardal et al.\ 2008; Mori \& Rich 2008) are correct, this would invalidate the Font et al.\ estimate of progenitor mass.

We do not yet have sufficient information to model the orbits of these five fainter streams in M31, and thus our estimates of the progenitor masses are slightly less secure than the GSS estimate in Font et al.\ (2006), with a linear dependence on the uncertain  $R_{peri}$. We place limits on $R_{peri}$ by assuming that we are seeing these streams at apocentre, and that orbits in cosmological simulations have average $R_{peri}/R_{apo}\sim 0.2 -- 0.25$ (van den Bosch et al.\ 1999; Ghigna et al.\ 1998; Benson 2005). 
These limits are listed in Table~1 along with the other model parameters (and taking for M31, $v_{circ}$=260~km/s).

The same procedure can also be applied to the Sagittarius stream (Sgr), taking parameters from 
Majewski et al.\ (2003), with  $R_{peri}=12$~kpc, w$_{FWHM}$=4~kpc, $v_{circ}=220$~kpc, we calculate a mass of 5.5$\times10^9$~M$_\odot$.
Majewski et al.\ (2003) derive a mass from kinematics of Sgr of 5.8$\times10^8$~M$_\odot$ with a M/L=25 (M$_V=-13.27$). 

The velocity dispersions of the streams can also be used to constrain the progenitor masses, although perhaps with even less accuracy.
Generally, the dispersion in stream debris 
should decrease over time (Helmi \& White 1999b). If these streams were very young, we wouldn't expect dynamical cooling to be 
signiÞcant yet. 
Tidal interactions with dark matter substructure in  
the halo may also not have had sufficient time to significantly heat the streams if they are relatively young (Ibata et al.\ 2002, Johnston et al. 2002).
Helmi \& White (1999b) suggest the velocity dispersion should 
vary most signiÞcantly in an oscillatory manner as a 
function of radial orbital phase.

Font et al.\ (2006) also estimate the progenitor mass of the GSS from a single 
$\sigma_{v_r}=15$~km/s lying  between apocentre and pericentre along the stream, with a lower limit of $10^{8}$ M$_\odot$, consistent with their estimate from structural/orbit properties. 
As predicted by Helmi \& White (1999),  the stream can become very cold in between the turning 
points with the velocity dispersion of the stream reaching values well below the central dispersion 
of the satellite, and as small as $\sigma/\sigma_0 \sim 0.5$.

In a similar manner, we can estimate the $M/L$ of the streams presented in this contribution.
For all the streams, we can apply the methodology of Font et al.\ (2006) directly as we have plausible kinematic detections in each case, along with reasonable constraints on the stream widths and morphologies.  We expect the velocity dispersion of the progenitor to be as large or larger than the intrinsic  value estimated for the stream and assume mass follows light (Richstone \& Tremaine 1986).
For \streamd, we have taken the combined $\sigma_{v_r}$ from both spectroscopic pointings (Table~1), assuming that the $\sim$-400~km/s stars are the most likely members of the structure.
For \streama, we assume the three halo stars at $-172$~km/s represent the stream (although as discussed, it is ambiguous whether it is a structure associated to M31 or M33).
The results of our mass estimates are presented in Table~1, and plotted on Fig.~12, where we highlight a factor of two uncertainty given that we are likely observing the streams at especially cold  points between turning.

However, we caution that these $M/L$ ratios have significant
uncertainties attached. The progenitor may not be completely disrupted, the velocity dispersion could be a very poor mass estimate (more typically a lower limit) depending on where in the orbit and evolution the stream is, 
and from the truncation of the image to the North, we already know 
the observed luminosity may not be representative for the entire streams.

The comparison of the streams with the Milky Way satellites and all published M31 satellites is presented in  Figure~12.  Stream M/L values are shown as bars connecting the structural mass estimate to the $\sigma_v$  estimate. 
The Mateo (1998) relation between M/L and the luminosity of a dwarf galaxies
in the Local Group has also been plotted on Figure~12. 
The relation can be understood physically as more massive dwarfs retaining more of their gas (and therefore arriving at $z=0$ with a lower M/L ratio), while lower mass halos more easily expel their gas and form smaller numbers of stars. 
All of our newly constrained M31 streams are in agreement with the relation within reasonable errors, which is somewhat surprising given our limited ability to constrain the masses of the streams.
Interestingly, the Sagittarius stream lies significantly above the relation over all plausible mass estimates.
Many of the new M31 dSphs, along with most MW dSphs fall on the relation, although some clearly appear to be outliers, as are many of the new faint Milky Way dSphs (Martin et al.\ 2007; Simon \& Geha 2007). 
This suggests that at the low mass end, a range of processes beyond simply the feedback in winds which explains galaxies as massive as our stream progenitors may truncate star formation  (e.g., Ricotti \& Gnedin 2005), or else the numerous model assumptions are failing when reaching these faint limits. 
It will be of interest to obtain improved constraints on these M31 streams and model their orbits, to see if progenitor mass estimates continue to keep them on the Mateo relation.

\section{Discussion}

Our spectroscopic survey of the M31 faint streams has yielded an encouraging initial census of the kinematics, however it is clear from our results that significant efforts with a 10~metre telescope are required to usefully constrain the kinematic properties of the streams for modeling.
At least in the case of the spatially overlapping \streamcr\ and \streamcp, it is
remarkable that we have been able to clearly distinguish these structures by kinematics. Indeed, it was not entirely clear from initial inspection of the imaging in Ibata et al.\ (2007) that there were two different streams in this vicinity at all. Only careful inspection of the imaging divided in slices of metallicity reveals two structures slightly offset spatially.


The cluster, \ec4, lies in a region where the metal-poor \streamcp\  has roughly 50\% the stellar density of the metal-rich \streamcr,
although our spectroscopy reveals that \ec4\ is likely related to \streamcp\
with \streamcr\ overlapping only in projection.
Could \streamcp\ actually be the debris from disrupted \ec4\ material? 
The integrated luminosity of \streamcp\ within the MegaCam survey is comparable to a small dwarf
galaxy like And\,XV or And\,XVI (Letarte et al.\ 2008), $M_V\sim-9.5$, which would suggest the baryonic matter mass loss of
\ec4\ ($M_V=-6.6$, Mackey et al.\ 2006) would be dramatically larger than its current {\it intact} mass.
No distortion of the \ec4\ isophotes is found in the HST imaging
of Mackey et al.\ (2006) (Tanvir et al.\ in prep), although the faintness of \ec4\ means this
is unlikely to be a good test of ongoing mass loss or tidal distortion.
We have however noted that the outer stars in \ec4\ show a statistically significant velocity shift from the inner \ec4\ stars, similar to the  \streamcp\ stars in the field surrounding \ec4. This could happen for instance if \ec4\ were disrupting in a stream along the line of sight.
Regardless, it is likely that \streamcp\ and \ec4\ are at least related by their
kinematics and metallicities, \ec4\ possibly representing an intact system carried along in the disrupted progenitor represented by \streamcp.
If \ec4\ is dark matter dominated, we have in fact detected the very first {\it sub-sub-halo} (i.e.\ a galaxy that was bound to
a satellite galaxy), possibly explaining its small (r$_c$=30pc) size.
From a $\Lambda$CDM cosmological point of view, the
LMC and SMC should  also have such sub-sub-halos with
$L\sim10^7$--$10^8$ L$_\odot$, but we find none (and here we cannot invoke tidal
disruption of these systems, because they seem to be falling in for the
first time).
Of course it also remains the possibility that \ec4\ has nothing to do with the \streamc\ structure at all.
Without precise distance information, it is difficult to rule this out completely.


We also ask whether  these stream structures could be related to the
giant southern stream imaged in Ibata et al.\ (2001, 2007)?
As we have noted, the metallicities of various streams 
all differ, only \streamcr\  and \streamb\ being
even close to the metallicity of the core region of the giant southern stream (Ibata et al.\ 2001, 2004, 2007; Guhathakurta et al.\ 2006).
However, in  
the CFHT-MegaCam survey, the outer region of the GSS are more metal-poor  
and have a metallicity that is similar to that of some of the more metal-poor streams.
Nonetheless, the general impression from the low dispersion of the streams in kinematics, their physical thickness, and varying (but narrow) metallicities makes it hard to reconcile with a scenario whereby these stream structures as an ensemble are related to the giant southern stream. 

Fardal (2008, and private communication) has modeled the stream resulting from
a GSS progenitor that is flattened and rotating like a disk, building on the constraints and models from
Fardal et al.\ (2007) and Gilbert et al.\ (2007).
Because it is on such a radial orbit, when it reaches pericenter very
close to M31, part of the progenitor could be on the opposite side of
M31 to the rest, depending on the orientation of the disk.  This means
that it starts orbiting M31 in the opposite direction to the rest
and leaves debris in different physical locations than the main stream.
The disk-like kinematics results in caustic structures that appear
similar to streams or arcs.
In this model, the new streams from Ibata et al. (2007), and herein, would be shells from this
counter-orbiting part of the stream.  Assuming a large metallicity
gradient in the progenitor, they would have a much lower metallicity
than the main stream since they come from an outer part of the progenitor.
Whether or not this specific model is correct,  the general idea 
that the progenitor was physically quite large, and passed 
extremely near the center of M31, means that debris could get thrown out 
in all directions. Given the metallicity structure of the progenitor, it is plausible that some of this debris could have distinct 
metallicities but be ultimately related to the same progenitor.
While somewhat implausible for the reasons stated above, it remains to be seen if the specific kinematics and metallicities of our new observations can be reproduced in such a model.

Whereas the photometric profiles could only remove Galactic contamination and stars belonging to stream substructures statistically, we can explicitly remove stars belonging to the streams (and the Milky Way) by their kinematics and assess the underlying M31 stellar halo density from 30--120~kpc on the minor axis.
While our resulting measurement has so few stars as to be highly uncertain statistically, it does reveal the general power of kinematic analysis of the M31 halo population.

It is remarkable the extent to which these kinematic substructures projected on the minor axis  dominate the halo star statistics in these fields. In the fields studied, they represent $\sim2/3$ of the candidate halo stars, revealing that the photometric minor axis profile from Irwin et al.\ (2005) and Ibata et al.\ (2007) is significantly flattened by such structures.
We are led to the likely conclusion that stellar halos are  
made up of multiple kinematically cold streams, perhaps even to the extent proposed by Bullock \& Johnston (2005) (see also Bell et al.\ 2008).





\section{Conclusions}

In conclusion, we have conducted a Keck/DEIMOS spectroscopic survey of five stellar streams, recently uncovered through deep imaging observations of the halo.

$\bullet$ We have uncovered a kinematic substructure at \vhel=-349.5$\pm1.8$~km/s from a spectroscopic
field lying in the Ibata et al.\ (2007) \streamc. The cold component has $\sigma_{v_r}=5.1\pm2.5$ and
a narrow range in [Fe/H]=-0.7$\pm$0.2, which we propose represents a metal-rich component, 
\streamcr.

$\bullet$ We have uncovered a second kinematic substructure in the same field as \streamcr\ at \vhel=-285.6$\pm1.2$~km/s with $\sigma_{v_r}=4.3^{+1.7}_{-1.4}$~km/s (non-zero at $>$3$\sigma$ confidence interval) and
a narrow range in [Fe/H]=-1.3$\pm$0.2, which we propose represents a metal-poor stream,
\streamcp. We demonstrated that this kinematic \streamcp\  has a counterpart in a spatially offset metal-poor region of \streamc\ in Ibata et al.\ (2007).

$\bullet$ We plausibly detect both \streamcr\  and \streamcp\ at a position $\sim$30kpc further north along the structure, with no detectable 
velocity gradiant for \streamcr, and a measured velocity gradient of $\sim$40~km/s for \streamcp.

$\bullet$ We were unable to identify kinematic substructure unambiguously 
associated to \streamd\  from our serendipitous spectroscopic pointing, 
however subsequent spectroscopy well centered in the \streamd\ identifies a likely cold kinematic 
structure which has a viable counterpart in the serendipitous pointing.
We propose a kinematic detection of \streamd\ at 
\vhel=-390.5~km/s with $\sigma_{v_r}=4.2$~km/s.

$\bullet$ Spectroscopy near the edges of \streama\ and \streamb\ suggest a likely kinematic detection for \streamb\ with
$v_{\rm hel}\sim-330$~km/s, $\sigma_{v, corr}\sim6.9$\,km/s,
and a kinematic detection of \streama\ at $v_{\rm hel}\sim-172$~km/s, $\sigma_{v_r}\sim12.5$\,km/s.
Neither spectroscopic pointing in these streams is ideally placed, and additional spectroscopic observations are well motivated to further constrain the kinematics of these structures.

$\bullet$ 
The extended cluster EC4 lies in the \streamc\ region, with 
kinematics (v$_{\rm hel}$=-285\,km/s) 
and metallicity ([Fe/H]=-1.4) which suggest it is related to the more metal-poor stream 
\streamcp. 
\ec4\ could be the progenitor of the metal-poor \streamcp\ (somewhat unlikely given the apparent stellar mass difference between the stream and \ec4), or it may simply be a structure carried along by the disrupted stream progenitor.
In this case, and if \ec4\ has a sizable dark matter component, we have in fact detected the very first {\it sub-sub-halo} 
(i.e.\ a galaxy that was bound to a satellite galaxy), possibly explaining its small (r$_c$=30~pc) size.

$\bullet$
By explicitly removing stars belonging to the streams by their kinematics we can assess the underlying M31 stellar halo density and metallicity on the minor axis. This contrasts the purely photometric approach where Galactic contamination and stars belonging to stream substructures can only be removed statistically.
Our resulting halo measurement has so few stars as to be highly uncertain statistically, however, it does reveal the general power of kinematic analysis of the halo population for future endeavors.
The fraction of background halo stars in these stream fields suggests the conclusion that stellar halos are  
largely made up of multiple kinematically cold streams.

\section{ACKNOWLEDGMENTS}
We thank an anonymous referee for a very careful reading and a helpful report. 
We thank Mark Fardal for helpful discussions on the giant southern stream.
SCC acknowledges NSERC and the Canadian Space Agency for support.
R.\ M.\ Rich and AK acknowledge support from AST-0309731.

\newcommand{\mnras}{MNRAS}
\newcommand{\pasa}{PASA}
\newcommand{\nat}{Nature}
\newcommand{\araa}{ARAA}
\newcommand{\aj}{AJ}
\newcommand{\apj}{ApJ}
\newcommand{\apjl}{ApJ}
\newcommand{\apjs}{ApJSupp}
\newcommand{\aap}{A\&A}
\newcommand{\aaps}{A\&ASupp}
\newcommand{\pasp}{PASP}

\begin{table*}
\begin{center}
\caption{Properties of DEIMOS fields in the five M31 stellar streams `A',`B',`Cp',`Cr',`D'.}
\label{tableSat}
\begin{tabular}{llcccccccc}
\hline
field & $\alpha$ (J2000), $\delta$ (J2000) & $v_{r, stream}$  & $\sigma_{stream }$ $^a$  & $<\FeH>$ $^b$ & w$_{stream}$ $^c$ & $R_{peri}$ $^d$ & $M_{stream}$ $^e$ & $L_{stream}$ $^f$\\
& {} & &  (km/s) &  (km/s) & (kpc) & (kpc) & $\times10^7$ M$_\odot$ & $\times10^6$ L$_\odot$ \\\hline
\hline
stream A (M8) & 01:14:01.37 +32:31:00.9 & -172.2 & 12.5 & $-1.3\pm0.3$ (-1.3) & 7.5 & 24 & 10/7 & 2.3\\
stream B (M6) & 01:32:14.64 +33:12:25.4 & -330.1 & 6.9 & $-0.8\pm0.2$ (-0.6) & 5.0 &16 & 7/2 & 10.0 \\
stream Cr (F25/F26) & 00:58:22.02 +38:04:05.9 & -349.5 & $5.1\pm2.5$ & $-0.7\pm0.2$ (-0.6)  & 6.8 & 12 & 30/1 & 12.6\\
stream Cp (F25/F26) & 00:58:22.02 +38:04:05.9& -287.3 & $4.3^{+1.7}_{-1.4}$ & $-1.3\pm0.2$ (-1.1) & 8.5 & 11 & 70/1 & 1.4 \\
stream Cr pos2 (F36) & 01:00:38.00 +38:45:37.0 & -350 & n/a & $-0.7\pm0.2$ (-0.6) & {n/a} \\
stream Cp pos2 (F36) & 01:00:38.00 +38:45:37.0 & -246 & n/a & $-1.2\pm0.2$ (-1.1) & {n/a} \\
stream D (F7) & 00:54:55.02 +39:43:55.3 & -390.5 & 4.2 & $-1.1\pm0.3$ (-1.2) & 8.2 & 6 & 213/1 & 9.5\\
stream D pos2 (F37) & 00:57:34.00 +39:49:12.0 & -390.5 & 4.2 & $-1.1\pm0.3$ (-1.2) & {n/a} \\
EC4 (F25/F26) & 00:58:15.50 +38:03:01.1    & -282.4 &$\sim10$ & -1.4$\pm$0.1/-1.6\,$^g$ & {n/a} \\
\hline
\end{tabular}
\end{center}
$^a$ Velocity dispersions, estimated through a maximum likelihood analysis taking into account the measurement errors in the velocities (or in the case of \streama, \streamb, \streamd, a subtraction in quadrature of the measurement error).\\
$^b$ For certain streams (notably \streama\ and \streamd), it is arguable that the [Fe/H]$_{\rm phot}$ 
measurements are estimated from sets of stars which may not actually be a kinematic
detection of "the stream". For this reason we also quote the statistical
[Fe/H]$_{\rm phot}$ estimate from Ibata et al.\ (2007) in all cases in brackets.
In particular the \streamd\ values are quoted for the combination of 6 stars in the two separated pointings along the stream, and therefore the numbers in the table are simply duplicated.\\
$^c$ Stream widths derived from Gaussian fits (quoted FWHM)  to the integrated profile in the region defined in Ibata et al.\ (2007) (their Fig.~31). At the distance of M31 (785~kpc -- McConnachie et al.\ 2005) 1~degree = 13~kpc.\\
$^d$ $R_{peri}$ estimated as 1/5 $D_{stream}$ as discussed in the text.\\
$^e$ Mass of stream width estimated from structural parameters (first entry) and from $\sigma_v$ (second entry).\\
$^f$ $L_{stream}$ from Ibata et al.\ (2007), except for \streamcp,\streamcr, which are discussed in the text.\\
$^g$ For EC4, the spectroscopic estimate of the [Fe/H] = -1.6, lying midway between the estimates derived using CFHT photometry with Girardi isochrones (-1.4) versus  and the HST photometry with Dartmouth isochrones (-1.8) as described in Collins et al.\ (in preparation).
\end{table*}

\begin{table*}
\begin{center}
\caption{Properties of candidate M31 halo stars in stream D$_1$ region (field F7).}
\label{tableSat}
\begin{tabular}{llcccccc}
\hline
$\alpha$ (J2000) & $\delta$ (J2000) & $v_r$ ($\kms$) & $v_{err}$ ($\kms$) & $\FeH_{phot}$ & $\FeH_{spec}$ & V-mag & I-mag\\\hline
00:54:18.02&+39:43:32.4&-449.7& 6.4& --& -3.12& 20.26& 19.38\\
00:54:26.77&+39:45:58.2&-150.3& 3.2& -1.78&-2.027& 22.31&21\\
00:54:31.71&+39:44:54.1&-346.7&9.96&-3&-1.662& 22.36&21.3\\
00:54:35.93&+39:43:59.2&-402.7& 11.05& -1.11&-3.295& 22.84& 21.32\\
00:54:39.48&+39:45:05.4&-169.4& 34.95& --&-2.181&21.7&21\\
00:54:40.93&+39:43:08.8&-268.4& 11.45& -0.73&-2.843&23.7& 21.29\\
00:54:47.03&+39:44:25.9&-370.8&8.01& -1.58&-3.454& 22.72& 21.44\\
00:54:50.30&+39:43:36.5&-237.3& 14.53& -1.43&-2.144& 22.61& 21.22\\
00:54:53.48&+39:43:11.3&-274.6&5.46&-3&-2.637& 22.66& 21.64\\
00:54:57.75&+39:43:32.5&-195.9& 49.42& -1.22&-3.394& 22.79& 20.76\\
00:55:00.87&+39:43:37.5&-254.2& 25.61& -1.12&-3.135&23& 20.82\\
00:55:10.40&+39:46:58.7&-162.1&2.73&--&-2.025& 21.46&20.7\\
00:55:12.94&+39:43:50.3&-151.5&2.44&-- &-2.427& 20.43& 19.62\\
00:55:13.27&+39:43:19.4&-211.5&7.26& --&-2.894& 23.69& 20.87\\
00:55:14.67&+39:44:52.5&-217.8& 38.62& -- &-3.182& 21.55&21\\
00:55:22.81&+39:45:25.5&-414.3&8.37& -1.18&-1.537& 22.86& 20.87\\
00:55:27.67&+39:43:25.5&-244.2&3.11& -- & -2.73& 21.42& 19.16\\
00:55:34.22&+39:42:51.8&-212.3& 12.14& --&-2.776& 23.52& 20.75\\
\hline
\end{tabular}
\end{center}
\end{table*}

\begin{table*}
\begin{center}
\caption{Properties of candidate M31 halo stars in stream D$_2$ region (field F37).}
\label{tableSat}
\begin{tabular}{llcccccc}
\hline
$\alpha$ (J2000) & $\delta$ (J2000) & $v_r$ ($\kms$) & $v_{err}$ ($\kms$) & $\FeH_{phot}$ & $\FeH_{spec}$ & V-mag & I-mag\\\hline
00:56:58.88&+39:50:08.3&-203.6&9.94& --&-1.384& 19.75& 18.35\\
00:57:02.80&+39:47:56.0&-150.9& 12.45& -1.19&0.4212&22.7& 21.18\\
00:57:06.19&+39:49:49.9&-178.6&6.47& -1.67&0.3611& 22.39& 21.05\\
00:57:09.96&+39:52:32.5&-366.9& 15.91&-1.9&-1.331& 24.11& 21.16\\
00:57:35.45&+39:49:42.4&-187.6& 24.21& -1.53& 2.214& 23.24&22.1\\
00:57:37.07&+39:47:50.2&-153.3& 4.8& --& -0.9179& 21.02& 19.73\\
00:57:40.78&+39:50:55.2&-264& 32.52& -0.54& -0.5198& 23.74& 21.93\\
00:57:41.37&+39:48:33.1&-197.6&3.56& -0.44& 1.305& 23.93& 22.07\\
00:57:43.61&+39:51:13.5&-192.6&8.33& -1.18& -0.3217& 22.77& 21.28\\
00:57:47.86&+39:50:35.0&-400.6& 43.33& -1.39&-0.605& 23.28&22.1\\
00:57:48.24&+39:48:36.6&-390.5& 10.44& -1.04& -0.3352&23& 21.01\\
00:57:50.48&+39:49:09.6&-169.2&4.03& -0.95&-1.017& 23.27&21\\
00:57:55.32&+39:51:25.7&-386.6&6.08& -0.74& -0.2843& 23.53& 21.08\\
00:57:56.65&+39:50:48.5&-385.6&8.55& --& -0.9452&21.1&20.1\\
00:57:57.41&+39:51:43.2&-233.7& 55.17& -0.84&-1.161& 23.42& 21.86\\
00:58:00.09&+39:48:33.5&-247.3&7.99& -0.84& -0.3931& 23.33& 21.33\\
00:58:06.18&+39:51:15.1&-258.6& 15.52& --&-1.195& 21.44& 20.61\\
\hline
\end{tabular}
\end{center}
\end{table*}

\begin{table*}
\begin{center}
\caption{Properties of candidate M31 halo stars in stream C$_1$ region (field F25/F26).}
\label{tableSat}
\begin{tabular}{llccccccc}
\hline
$\alpha$ (J2000) & $\delta$ (J2000) & $v_r$ ($\kms$) & $v_{err}$ ($\kms$) & $\FeH_{phot}$ & $\FeH_{spec}$ & V-mag & I-mag & $D_{EC4}\,^a$\\\hline
00:57:46.69&+38:11:43.0&-349& 6.3& -1.22& -1.28& 22.49& 20.47& 10.38\\
00:57:48.22&+38:11:48.6&-343.6&8.38& -0.89& -4.83& 23.29& 21.72&10.3\\
00:57:51.02&+38:11:25.4&-193.3&5.69& -0.55& -1.75& 23.74& 21.83& 9.68\\
00:57:52.27&+38:08:34.6&-289.3&8.68& -1.01& -1.76& 23.24& 21.82& 7.19\\
00:57:54.36&+38:12:26.8&-286.9&5.21& -1.29& -0.63& 22.57& 20.98&10.3\\
00:57:56.30&+38:06:18.1&-356.5& 6.7& -0.62&2.75&23.7& 21.61& 5.00\\
00:58:05.23&+38:08:41.4&-223&2.22&-0.4&-1.8& 24.27& 21.48&6.02\\
00:58:12.75&+38:05:24.6&-341.8& 14.37&-0.7& -0.97& 23.58& 21.51& 2.45\\
00:58:13.99&+38:06:18.1&-295.1&5.96& -1.26& -1.09& 22.99& 21.66& 3.30\\
00:58:14.37&+38:03:00.4&-296&5.29& -1.48& -1.84& 22.91& 21.65&0.22\\
00:58:14.74&+38:03:00.8&-277.3&9.56& -1.49& -5.62& 22.77& 21.47&0.15\\
00:58:15.24&+38:03:01.3&-293.4& 11.55& -1.39& -1.89&21.7& 20.99& 0.05\\
00:58:15.32&+38:03:09.7&-371.5&10.9& -2.17&0.63& 22.65& 21.57&0.15\\
00:58:15.47&+38:02:58.9&-267.4&8.42& -1.45& -0.53& 22.41& 20.92& 0.04\\
00:58:15.94&+38:04:34.9&-153.3& 40.28& -2.45&3.38& 22.78& 21.73& 1.57\\
00:58:15.99&+38:02:56.1&-264&7.74& -1.21& -0.69& 22.75&21.3&0.13\\
00:58:16.00&+38:02:22.5&-280.3&2.83& -1.41& -1.38& 22.06& 20.41&0.65\\
00:58:17.12&+38:02:54.1&-281& 2.9& -1.36& -0.65& 22.49& 20.86&0.34\\
00:58:17.16&+38:02:49.6&-290&2.57& -1.16& -1.12& 22.68& 21.09&0.38\\
00:58:18.62&+38:00:16.7&-530.7&9.81&--&0.73& 24.18&20.6& 2.81\\
00:58:19.68&+38:06:42.4&-577.2& 11.45& -0.77&-2.2&23.5& 21.11& 3.78\\
00:58:20.25&+37:59:46.6&-183.4& 19.86& -0.56& -0.96& 23.84& 21.65& 3.37\\
00:58:21.87&+38:00:13.9&-344.7&4.98& -1.04& -0.69& 22.87& 21.06& 3.06\\
00:58:22.02&+38:04:05.9&-324.3& 20.49& -0.84& 2.1& 23.39& 21.84& 1.68\\
00:58:23.15&+37:58:40.9&-159.6&8.61& -2.25& -1.27& 22.26&21.1& 4.59\\
00:58:24.32&+38:04:29.9&-284.3& 2.6&--& -3.06& 20.43& 19.71&2.28\\
00:58:25.71&+38:03:12.2&-348.1& 11.12& -0.65&-0.1& 23.74& 21.53& 2.02\\
00:58:29.49&+38:00:03.6&-360.7& 19.15& -0.82&-0.2& 23.37& 21.75& 4.04\\
00:58:32.37&+37:59:42.3&-293.3&9.84& -1.09& -0.98& 23.14& 21.75& 4.69\\
00:58:34.09&+37:56:39.3&-157.9&2.34&--& -2.22& 21.33& 20.81& 7.34\\
00:58:41.07&+37:55:54.7&-461.1& 8.3& -0.37& -1.34& 24.16& 21.91& 8.71\\
00:58:42.74&+38:00:24.9&-220.8&3.74& -0.47& -1.52& 23.98& 21.44& 5.96\\
00:58:43.46&+38:00:22.4&-375.8&8.44& -1.18&-0.7& 22.84& 20.86& 6.11\\
00:58:44.18&+38:00:08.9&-338&7.35& -1.02& -1.35& 23.02& 21.46& 6.33\\
00:58:47.15&+37:55:52.5&-361.7&5.72& -0.86& -1.55& 23.38& 21.83& 9.48\\
\hline
\end{tabular}
\end{center}
$^a$ Distance from the center of EC4 in arcmin. At the distance of EC4, 13~kpc = 1deg.
\end{table*}

\begin{table*}
\begin{center}
\caption{Properties of candidate M31 halo stars in stream C$_2$ region (field F36).}
\label{tableSat}
\begin{tabular}{llcccccc}
\hline
$\alpha$ (J2000) & $\delta$ (J2000) & $v_r$ ($\kms$) & $v_{err}$ ($\kms$) & $\FeH_{phot}$ & $\FeH_{spec}$ & V-mag & I-mag\\\hline
01:00:01.15&+38:44:42.5&-150.3& 71.98&-1.3&-2.132& 23.32& 22.14\\
01:00:02.17&+38:47:18.9&-212.4&5.96& --& -2.05& 19.74& 18.54\\
01:00:02.65&+38:46:11.4&-164.5& 21.39& --&-2.101& 21.39& 19.12\\
01:00:03.35&+38:46:38.3&-308&8.92& -1.02&-1.052&22.9& 21.08\\
01:00:05.16&+38:44:44.3&-237.3& 13.17& -1.37&-2.313& 22.51& 21.04\\
01:00:05.98&+38:45:08.8&-184.2& 12.84& -1.88&-2.384&22.4& 21.18\\
01:00:07.06&+38:45:41.5&-209.1&5.81& -1.07&-1.145& 23.26& 21.91\\
01:00:10.97&+38:48:13.5&-348.6&6.91& -1.21& -0.940& 22.65& 21.13\\
01:00:12.06&+38:47:41.6&-190.8& 53.77& -1.33&-1.991& 22.52& 21.01\\
01:00:15.76&+38:46:00.4&-448.9&9.47& -0.88&-1.234& 23.47& 22.03\\
01:00:17.03&+38:45:40.0&-350.3& 14.61& -0.78&-1.159& 23.37& 21.54\\
01:00:20.95&+38:45:43.5&-369.6&9.73& -0.73&-1.252& 23.43& 21.61\\
01:00:23.72&+38:45:44.2&-630.9&9.66& --&-2.244&21.7&21.1\\
01:00:25.69&+38:47:17.5&-504.2& 13.56& -0.66& -1.42& 23.58& 21.87\\
01:00:31.57&+38:44:49.2&-393.2&8.67& -0.96&-1.101& 22.98& 21.18\\
01:00:33.31&+38:47:36.2&-239.1&41.9& -1.16& -0.6722& 23.24& 21.94\\
01:00:42.01&+38:44:57.3&-229.1&7.31&-1& -0.1138& 23.15& 20.96\\
01:00:43.36&+38:46:32.3&-497.4& 15.02& --&-1.931& 24.45& 21.18\\
01:00:45.42&+38:47:38.4&-302.4& 15.83& -0.66&-1.144& 23.59& 21.88\\
01:00:46.31&+38:46:36.4&-382.4& 16.62& -0.63& -0.7123& 23.62& 21.65\\
01:00:47.34&+38:46:13.4&-493& 31.84& -1.08&-2.085&23.3& 21.98\\
01:00:47.67&+38:46:53.0&-341.4& 13.17& -0.77& -0.8154& 23.39&21.5\\
01:00:53.79&+38:47:58.1&-166.8&8.33& --&-2.543& 20.18& 19.61\\
01:00:54.98&+38:44:22.9&-419.9& 38.68& -0.92&-1.229& 23.45& 22.05\\
01:00:56.02&+38:45:19.6&-253.4& 25.28& --&-1.675& 21.32& 20.57\\
01:00:58.50&+38:46:59.7&-317.4& 10.78& --&-2.282&20.9& 20.08\\
01:01:01.43&+38:46:30.6&-319.6& 17.02& -1.22&-1.312& 22.87& 21.47\\
01:01:04.01&+38:48:43.0&-341.3& 24.47& --&-2.343& 21.33& 20.37\\
01:01:15.28&+38:47:02.7&-175.8&3.57& --&-2.081& 20.32& 18.28\\
\hline
\end{tabular}
\end{center}
\end{table*}

\begin{table*}
\begin{center}
\caption{Properties of candidate M31 halo stars in stream B region (field M6).}
\label{tableSat}
\begin{tabular}{llcccccc}
\hline
$\alpha$ (J2000) & $\delta$ (J2000) & $v_r$ ($\kms$) & $v_{err}$ ($\kms$) & $\FeH_{phot}$ & $\FeH_{spec}$ & V-mag & I-mag\\\hline
{01:08:31.0} &   {37:30:21.6} &      -335.4   &     10.1    &   -0.91   &   -1.28    &   22.51&   20.78\\
{01:08:33.9} &   {37:32:47.0} &      -292.3   &    6.571   &    -10.26&      -2.74 &    22.49&      21.6\\
{01:08:34.8} &   {37:29:19.1} &      -435.1   &    10.48   &      -0.2   &    --   &    23.94&   21.94\\
{01:08:36.4} &   {37:34:00.4} &      -354.8   &    10.82   &     -0.03  &    -2.25   &    24.38&      22\\
{01:09:36.4} &   {37:52:43.4} &      -317.3   &     7.16   &     -1.04    &  -0.97  &     22.74&    21.25\\
{01:09:36.4} &   {37:52:57.4} &      -327.3   &    5.347  &      -0.67   &    --  &    23.08&   21.37\\
{01:08:36.5} &   {37:25:16.6} &      -219.9   &    14.38   &     -0.36  &    -2.96    &   23.62&      21.23\\
{01:09:42.4} &   {37:47:47.7} &      -369.9   &    13.09  &       -2.1    &  -1.61  &     22.26&   20.99\\
{01:09:43.1} &   {37:41:33.4} &      -152.8   &    9.844   &     -0.58   &    -2.1  &     23.15&   21.15\\
{01:09:48.0} &   {37:51:28.1} &      -307.8   &    11.37   &     -0.47   &   -0.18 &    23.61&     20.89\\
{01:09:50.9} &   {37:43:20.3} &      -274.1   &    5.322   &     -1.17   &    --  &    23.39&     22.1\\
{01:09:53.9} &   {37:52:18.1} &      -385.5   &     13.3   &     -0.64    &  -1.52  &     23.06&   21.24\\
\hline
\end{tabular}
\end{center}
\end{table*}

\begin{table*}
\begin{center}
\caption{Properties of candidate M31 halo stars in stream A region (field M8).}
\label{tableSat}
\begin{tabular}{llcccccc}
\hline
$\alpha$ (J2000) & $\delta$ (J2000) & $v_r$ ($\kms$) & $v_{err}$ ($\kms$) & $\FeH_{phot}$ & $\FeH_{spec}$ & V-mag & I-mag\\\hline
{01:18:11.4} &  {36:12:51.4} &    -308.6  &     10.49 &     -1.58  &        -2.0 
&      22.41&    21.08 \\
{01:18:30.2} &  {36:22:24.7} &    -178.1  &     9.126 &      -1.5  &        --
&      22.71&    21.4 \\
{01:18:31.2} &  {36:17:09.8} &    -178.4  &     6.106 &      -1.12 &      -1.88
&      22.38&    20.83 \\
{01:18:32.0} &  {36:13:03.5} &    -153.2  &     5.416 &      -1.41 &      -1.55
&      22.16&    20.7 \\
\hline
\end{tabular}
\end{center}
\end{table*}

\end{document}